\documentclass{PoS}

\title{New Theoretical Results on \\ Event-by-Event Fluctuations}

\ShortTitle{Event-by-Event Fluctuations}

\author{\speaker{Mark I. Gorenstein}
\\
        Bogolyubov Institute for Theoretical Physics, Kiev, Ukraine\\
        E-mail: \email{goren@bitp.kiev.ua}}

\abstract{Several theoretical results concerning  event-by-event
fluctuations are discussed: \\
(1) a role of the global conservation laws and concept of statistical ensembles;\\
(2) strongly intensive measures
for  physical systems with volume fluctuations;\\
(3) identity method
for chemical fluctuations in a case of incomplete particle identification;\\
(4) the example of  particle number fluctuations in a vicinity of the critical point.}

\FullConference{9th International Workshop on Critical Point and Onset of Deconfinement - CPOD2014,\\
		17-21 November 2014\\
		ZiF (Center of Interdisciplinary Research)\newpage, University of Bielefeld, Germany}

\usepackage{amsmath}
\usepackage{amssymb}

\usepackage{epsfig}

\newcommand{\eq}[1]{\begin{align} #1 \end{align}}

\begin{document}

\section{Introduction}
The study of event-by-event (e-by-e) fluctuations in high-energy
nucleus-nucleus (A+A) collisions opens new possibilities to
investigate properties of strongly interacting matter
(see, e.g., Refs.~\cite{Koch:2008ia} and \cite{GGS-2014} and references therein).
Specific fluctuations can signal the onset of deconfinement when the
collision energy becomes sufficiently high to create the
quark-gluon plasma (QGP) at the initial stage of A+A
collision~\cite{dyn-fluc1,dyn-fluc2}.  By measuring the fluctuations, one may also observe
effects caused by the dynamical instabilities when the expanding
system goes through the 1$^{st}$ order transition line between the
QGP and the hadron resonance gas~\cite{fluc2}. Furthermore,
the critical point (CP) of strongly interacting matter
may be signaled by  characteristic
fluctuation pattern~\cite{fluc3,Koch:2005vg,Koch:2005pk}.
Therefore, e-by-e fluctuations are an important tool for the study of properties of the
onset of deconfinement and the search for the
CP of strongly interacting matter.
However, mostly due to the incomplete acceptance of detectors,
difficulties to control e-by-e the number of interacting
nucleons, and also not well
adapted data analysis tools, the results on e-by-e fluctuations are not
yet mature. Even the simplest tests of statistical and dynamical models
at the level of fluctuations are still missing.

In this presentation the theoretical progress in several areas related to
the study of e-by-e
fluctuations is reported.
A role of the global conservation laws is discussed in Sec.~\ref{s-GCL}.
The strongly intensive measures of e-by-e fluctuations
are introduced in Sec.~\ref{s-SIQ}. They give a possibility to study e-by-e fluctuations
in a physical system when its average size and size fluctuations
can not be controlled experimentally.  In Sec.~\ref{s-IPI} a novel procedure, the identity method,
is described for analyzing fluctuations of identified hadrons under typical experimental
conditions of incomplete particle identification.
Finally, in Sec.~\ref{s-CP} using the van der Waals
equation of state adopted to the grand canonical ensemble formulation
we discuss particle number fluctuations in a vicinity of the CP.
Most part of our discussion concerns the e-by-e fluctuations of hadron
multiplicities. However, many of our physical conclusions can be applied to more
general cases.

\section{Global Conservations Laws}\label{s-GCL}
In this section we illustrate the role of global conservation laws
in calculating of e-by-e fluctuations within statistical mechanics.
Successful applications of the statistical model to description
of mean hadron multiplicities in high energy collisions (see,
e.g., Refs.~\cite{stat1,stat2,stat3} and references therein) has
stimulated investigations of properties of the statistical ensembles.
Whenever possible, one prefers to use the grand canonical ensemble
(GCE) due to its mathematical convenience. The canonical ensemble
(CE) should be applied~\cite{CE1,CE2} when the number of carriers of
conserved charges is small (of the order of 1), such as strange
hadrons~\cite{strange}, antibaryons~\cite{antibaryons}, or
charmed hadrons \cite{charm}. The micro-canonical ensemble
(MCE) has been used~\cite{MCE1,MCE2,MCE3} to describe small systems with
fixed energy, e.g., mean hadron multiplicities in proton-antiproton
annihilation at rest. In all these cases, calculations performed
in different statistical ensembles yield different results. This
happens because the systems are `small' and they are `far away'
from the thermodynamic limit (TL). The multiplicities of
hadrons produced in relativistic heavy ion collisions are typically much larger than 1.
Thus, their mean values obtained within
GCE, CE, and MCE approach each other. One refers here to the
thermodynamical equivalence of statistical ensembles in the TL and uses the
GCE, as a most convenient one, for calculating the hadron yields.


A statistical system is characterized by the extensive quantities:
volume $V$, energy $E$, and conserved charge(s)\footnote{
These
conserved charges are usually the net baryon number, strangeness,
and electric charge. In non-relativistic statistical mechanics
the number of particles plays the role of a conserved `charge'.}
$Q$. The MCE is defined by the postulate that all micro-states
with given $V$, $E$, and $Q$ have equal probabilities of being
realized. This is the basic postulate of statistical
mechanics. The MCE partition function just calculates the number
of microscopic states with given fixed $(V,E,Q)$ values.
In the CE the energy exchange between the considered system and
`infinite thermal bath' is assumed. Consequently, a new parameter,
temperature $T$, is introduced.
To define the GCE, one makes a similar construction for the
conserved charge $Q$:  an `infinite chemical bath' and the
chemical potential $\mu$ are introduced.
The CE introduces the energy fluctuations. In the GCE, there are
additionally the charge fluctuations.
Therefore, the global conservation laws of $E$ and $Q$ are treated
in different ways: in the MCE both $E$ and $Q$ are fixed in each microscopic state,
whereas only average value of $E$ is fixed in the CE, and average values
of $E$ and $Q$ in the GCE.

The MCE, CE, and GCE are the most familiar statistical ensembles. In
several textbooks (see, e.g., Ref.~\cite{RR,Tolpygo}), the
pressure (or isobaric) ensemble has been also discussed.
An `infinite bath of the fixed external pressure' $p$ is then
introduced. This leads to volume fluctuations around the
average value (see Ref.~\cite{G-2008}).
In general, there are 3 pairs
of variables -- $(V,p),~ (E,T),~ (Q,\mu)$~ -- and, thus, the 8
statistical ensembles\footnote{For several conserved charges
$\{Q_i\}$ the number of possible ensembles is larger, as each
charge can be treated either canonically or grand canonically.}
can be constructed.

Measurements of  hadron multiplicity distributions in
A+A collisions, open a new
field of applications of the statistical models. The particle
multiplicity  fluctuations are usually quantified by a ratio of
the variance to mean value, the
scaled variance,
\eq{\label{scaled-var}
\omega[N]~\equiv~\frac{\langle N^2\rangle ~-~\langle N\rangle^2}{\langle N\rangle}~,
}
and are a subject of current experimental
activities. In statistical models there is a qualitative
difference in properties of a mean multiplicity and a scaled
variance of multiplicity distribution. It was recently found
\cite{fluc-1,fluc-2,fluc-3,fluc-4,fluc-5,fluc-6,mce-1,mce-2,res,exp,clt,acc}
that even in the TL
corresponding results for the scaled variance are different in
different ensembles. Hence the equivalence of ensembles holds for
mean values in the TL, but does not extend to fluctuations.
Several examples below illustrate
this statement.
\subsection{Canonical Ensemble}\label{ss-ce}
Let us consider a system which consists of one sort of
positively +1 and negatively -1 charged particles (e.g., $\;\pi^+\;$ and
$\;\pi^-\;$  mesons) with total charge equal to zero $\;Q=0\;$.
For the relativistic Boltzmann ideal gas
in the volume $V$  at temperature
$T$ the GCE partition function reads:
\begin{align}\label{Zgce}
Z_{\rm gce}(V,T) \;=\;
 \sum_{N_+=0}^{\infty}\sum_{N_-=0}^{\infty}\;
 \frac{(\lambda_+z)^{N_+}}{N_+!}\;\frac{(\lambda_-z)^{N_-}}{N_-!}
\;=\; \exp\left(\lambda_{+}z~+~\lambda_{-}z\right)~=~\exp(2z)~.
\end{align}
In Eq.~(\ref{Zgce}) $\;z\;$ is a single particle partition function
\begin{align}\label{z}
z\;=\; \frac{V}{2\pi^2}
       \int_{0}^{\infty}k^{2} dk\;
       \exp\left[-~\frac{(k^{2}+m^{2})^{1/2}}{T}\right]
        \;=\; \frac{V}{2\pi^2} \;\;
       T\,m^2\,K_2\left(\frac{m}{T}\right)~,
\end{align}
where $m$ is a particle mass, and $\;K_2\;$ is the modified Hankel
function.
Parameters $\;\lambda_+\;$ and $\;\lambda_-\;$ are auxiliary
parameters introduced  in order to calculate  mean numbers and
fluctuations of the positively and negatively charged particles.
They are set to one in the
final formulas. The chemical
potential  equals zero to satisfy the condition
$\langle Q\rangle_{\rm gce}=0$.

\begin{figure}
\includegraphics[width=.52\textwidth]{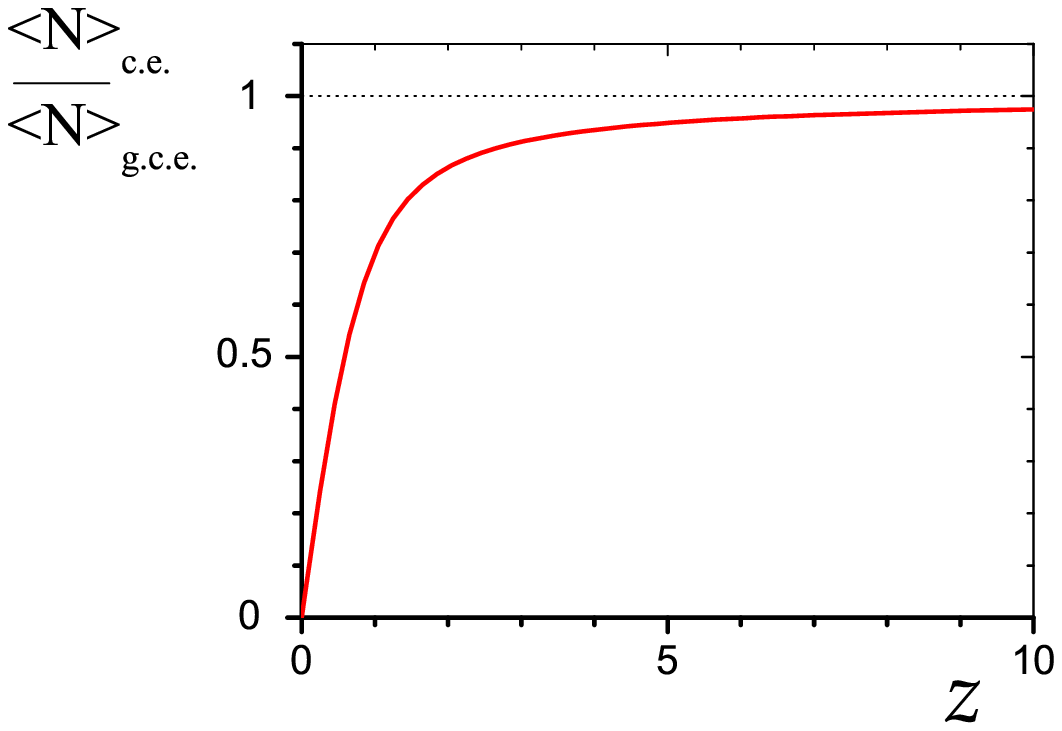}
\includegraphics[width=.46\textwidth]{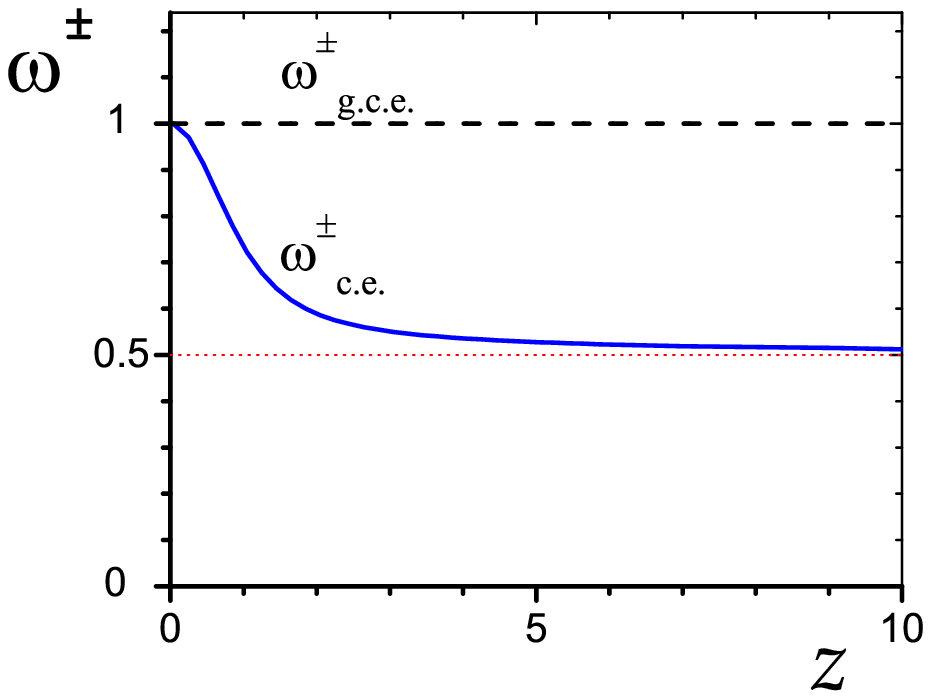}
\caption{ {\it Left:}
 The ratio of $\langle N_{\pm}\rangle_{\rm ce}$ (\protect\ref{ce-average})
 to $\langle N_{\pm} \rangle _{\rm gce}$ (\protect\ref{gce-average})  as a
 function of $z$. {\it Right:} The scaled variances of $N_{\pm}$
calculated within the GCE, $\omega^{\pm}_{\rm gce}=1$
(\protect\ref{omega-gce}), and CE, $\omega^{\pm}_{c.e.}$ (\protect\ref{omega-ce}).
}\label{fig-ce}
\end{figure}

The CE  partition function
is obtained by an explicit introduction of the charge
conservation constrain, $Q=N_+ - N_- = 0$, for each microscopic state
of the system and it reads:
\begin{align}\label{Zce}
Z_{\rm ce}(V,T)
 &\;=\;
 \sum_{N_+=0}^{\infty}\sum_{N_-=0}^{\infty}\;
 \frac{(\lambda_+ z)^{N_+}}{N_+!}\;\frac{(\lambda_- z)^{N_-}}{N_-!}
 \;\delta (N_+-N_-) \;=\;
  \\
 &\;=\;
 \frac{1}{2\pi}\int_0^{2\pi}d\phi\;\;
   \exp\left[ z\;(\lambda_+\;e^{i\phi}
                   \;+\; \lambda_-\;e^{-i\phi})\right]
  \;=\; I_0(2z)\;, \nonumber
\end{align}
where  the integral representations of the
$\delta$-Kronecker symbol
and the modified Bessel function $I_0$  were used.
The average
number of $N_{+}$ and $N_{-}$ can be calculated as \cite{Raf}:
\begin{align} \label{gce-average}
& \langle N_{\pm}\rangle_{\rm gce} \;=\;
 \left( \frac{\partial}{\partial\lambda_{\pm}}\ln Z_{\rm gce}
 \right)_{\lambda_{\pm}\;=\;1} =\; z~. \\
%
\label{ce-average}
& \langle N_{\pm}\rangle_{\rm ce} \;=\;
 \left( \frac{\partial}{\partial\lambda_{\pm}}\ln Z_{\rm ce}
 \right)_{\lambda_{\pm}\;=\;1}
\;=\; z\;\frac{I_{1}(2z)}{I_0(2z)}\;.
\end{align}
The exact charge conservation leads to the CE suppression,
$I_{1}(2z)/I_{0}(2z)<1$, of the charged particle multiplicities relative to
the results for the GCE (\ref{gce-average}).
The ratio of $\langle N_{\pm} \rangle$ calculated in the
CE and GCE is plotted as a function of $z$ in Fig.~\ref{fig-ce} {\it left}.

The corresponding scaled variances are \cite{fluc-1}:
\begin{align}\label{omega-gce}
\omega_{\rm gce}^{\pm}
 & \;=\; \frac{\langle N_{\pm}^2\rangle_{\rm gce}
        \;-\; \langle N_{\pm}\rangle_{\rm gce}^2}
{\langle N_{\pm}\rangle_{\rm gce}}
   \;=\; 1~,\\
\omega_{\rm ce}^{\pm}
 & \;=\; \frac{\langle N_{\pm}^2\rangle_{\rm ce}
        \;-\; \langle N_{\pm}\rangle_{\rm ce}^2}{\langle N_{\pm}\rangle_{\rm ce}}
   \;=\; 1 \;-\; z\left[\,\frac{I_1(2z)}{I_0(2z)}
           \;-\; \frac{I_2(2z)}{I_1(2z)}\,\right]~.\label{omega-ce}
\end{align}

In the large volume limit ($V\rightarrow \infty$ corresponds also to
$z\rightarrow \infty$) one can use an  asymptotic expansion
of the modified Bessel function,
\begin{align}\label{Bessel-0lim}
\lim_{z\to\infty}I_n(2z)\;=\;\frac{\exp(2z)}{\sqrt{4\pi z}}\;\left[1~-~
\frac{4n^{2}-1}{16z}~+~O\left(\frac{1}{z^{2}}\right)\right]~,
\end{align}
%
and obtains:
\begin{align}\label{ce1}
& \langle N_{\pm} \rangle _{\rm ce}~\cong~ \langle N_{\pm} \rangle _{g.c.e} =
z~, \\
& \label{omega-ce1}
        \omega_{\rm ce}^{\pm}
  \; \cong \; \frac{1}{2} ~ + ~ \frac{1}{8z}~\cong ~\frac{1}{2}~=~\frac{1}{2}
  ~\omega_{\rm gce}^{{\pm}}\;.
\end{align}
The dependence of the scaled variance calculated within
the CE and GCE on $z$ is shown in Fig.~\ref{fig-ce} {\it right}.
The scaled variance shows a very different behavior than the mean
multiplicity:
in the large $z$ limit the mean multiplicity
ratio approaches one and the scaled variance ratio 1/2.
Thus, in the case of $N_{\pm}$ fluctuations the CE  and
GCE are not equivalent.

\subsection {Micro-Canonical Ensemble}\label{ss-mce}

Our second example is the ideal gas of massless neutral Boltzmann particles.
We consider the same volume and energy in the MCE and GCE and
compare these to formulations. The fixed MCE energy $E$ and  the
mean GCE energy $\langle E\rangle_{\rm gce}$ are  connected
via equation (the degeneracy
factor is $g=1$):
\eq{\label{E}
E~=~\langle E\rangle_{\rm gce}~=~\frac{3}{\pi^2}~V~T^4~.
}
%
The mean multiplicity $\langle N\rangle_{\rm mce}$ in the MCE
is approximately equal to the  GCE value:
\eq{\label{N-ov}
\langle N\rangle_{\rm mce}~\cong~\langle N\rangle_{\rm gce}~=~\frac{1}{\pi^2}~V~T^3
~\equiv~  \overline{N} ~.
}
The approximation $\langle N\rangle_{\rm mce} \cong \overline{N}$ is
valid for $\overline{N}\gg 1$ and reflects the thermodynamic
equivalence of the MCE and GCE. The scaled variances for the multiplicity
fluctuations are, however, different in the GCE and  MCE \cite{mce-1}:
\eq{\label{om-gce}
 \omega_{\rm gce}~&=~\frac{\langle N^2\rangle_{\rm gce}~
-~\langle N\rangle_{\rm gce}^2}{\langle N\rangle_{\rm gce}}~=~1~,\\
 \omega_{\rm mce}~&=~\frac{\langle N^2\rangle_{\rm mce}~-~
\langle N\rangle_{\rm mce}^2}{\langle N\rangle_{\rm mce}}~=~\frac{1}{4}~.\label{om-mce}
}
 Thus, despite of thermodynamic
equivalence of the MCE and GCE the value of $\omega_{\rm mce}$ is four
times smaller than the scaled variance of the GCE (Poisson)
distribution, $\omega_{\rm gce} = 1$.
\begin{figure}[ht!]
\epsfig{file=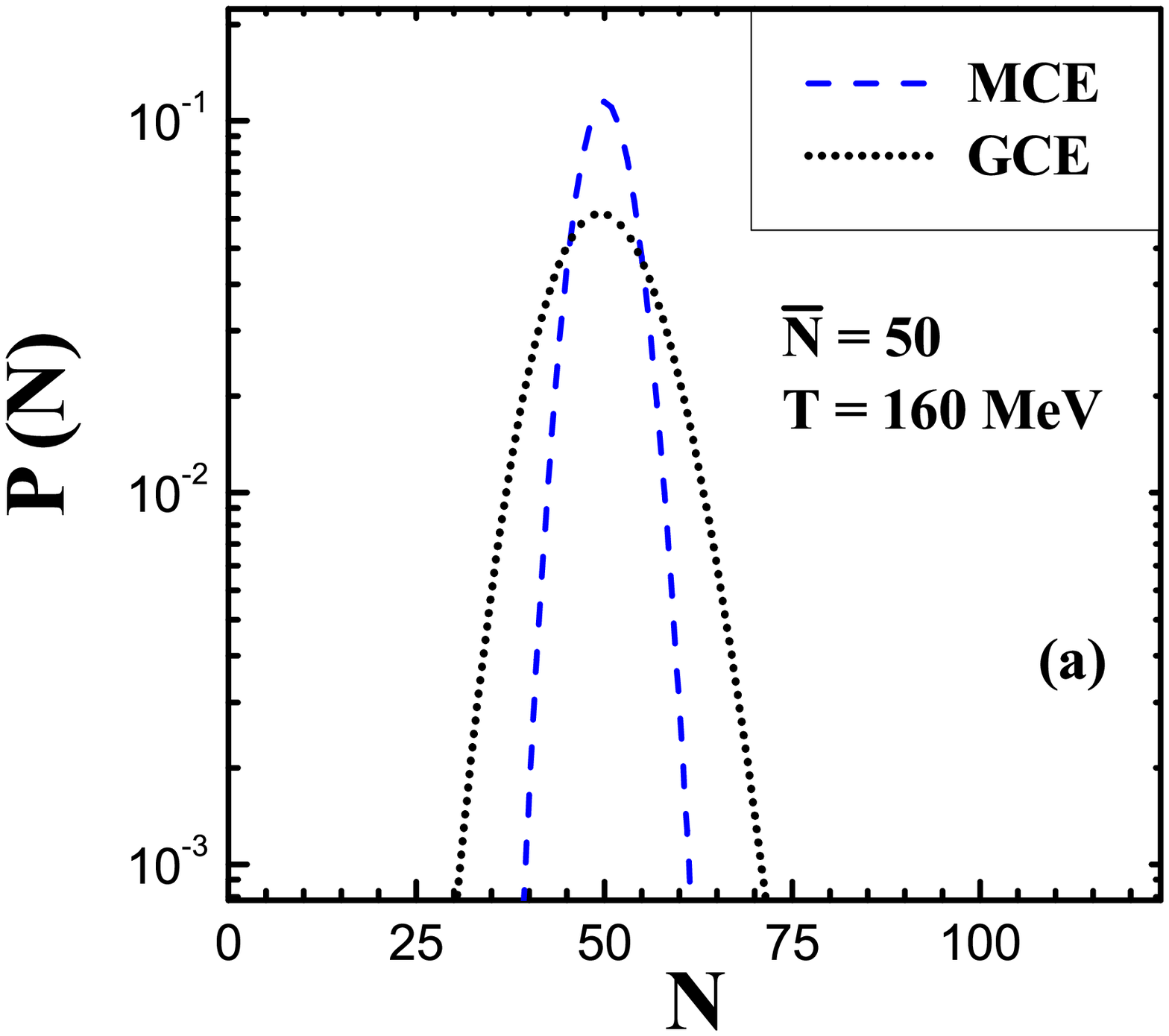,width=0.49\textwidth}\;\;
\epsfig{file=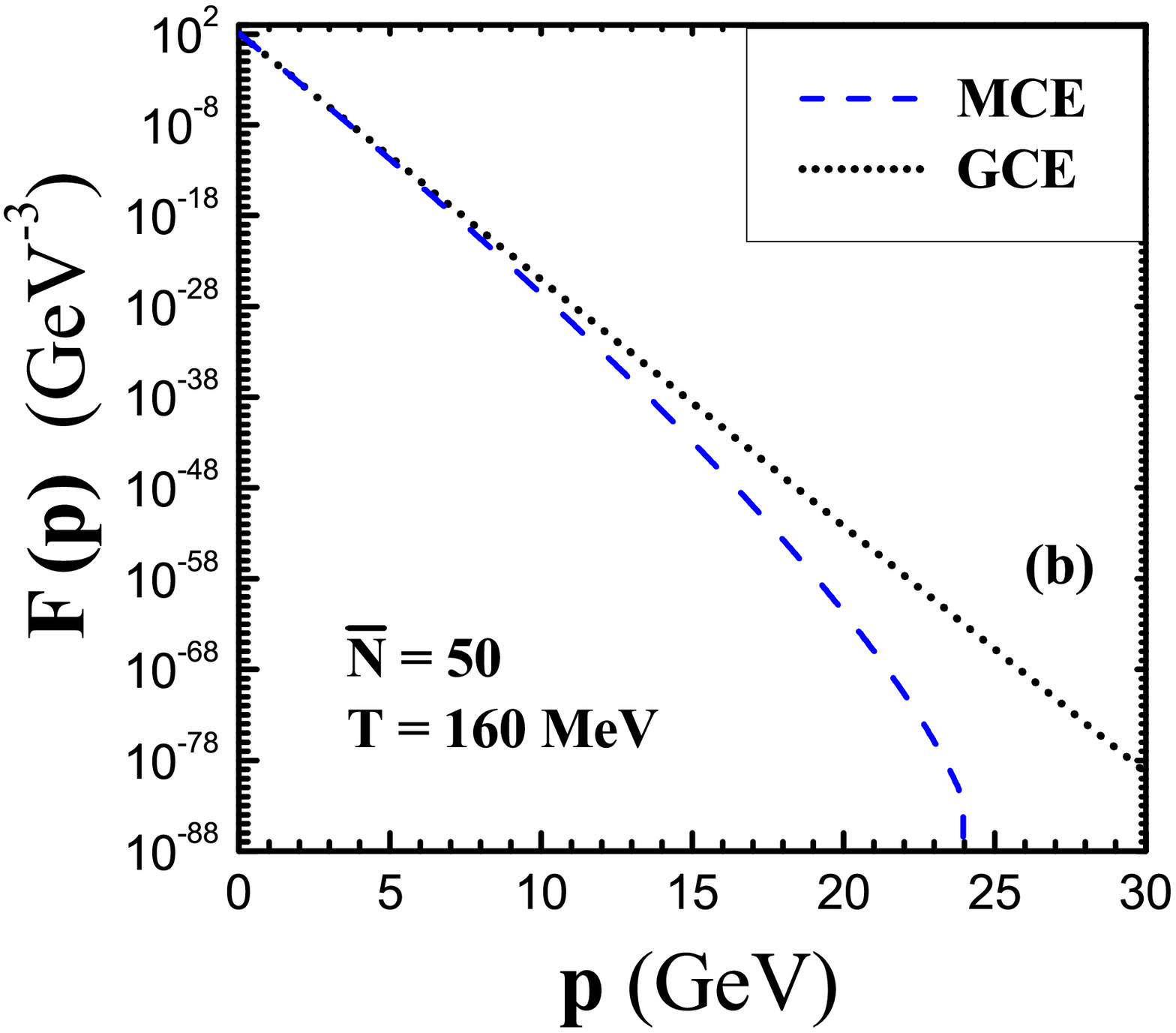,width=0.49\textwidth}
\caption{ {\bf (a):} The multiplicity distribution $P(N)$
of massless neutral particles in the MCE, dashed
line, and the GCE (Poisson) distribution, dotted line,
for the same mean multiplicity, $\overline{N}=50$.
{\bf (b):} The momentum spectrum of massless neutral particles
calculated within the MCE, dashed line, and the GCE
(\protect\ref{Boltz}), dotted line. The temperature is fixed as $T=160$~MeV and the system energy is
$E=3\overline{N}T=24$ GeV for both plots.
}  \label{fig-p-distr}
\end{figure}
%
%
%
%
Figure~\ref{fig-p-distr}(a) shows a comparison of the MCE and GCE
results for the multiplicity distributions $P(N)$. Different values of the scaled variances
(\ref{om-gce}) and (\ref{om-mce}) signify the different widths of $P(N)$
distributions  in the GCE abd MCE clearly seen in Fig.~\ref{fig-p-distr}.

Figure~\ref{fig-p-distr}(b) shows the single particle momentum spectra $F(p)$
in the GCE and MCE
($\overline{N}=50$ and  $T=160$~MeV).
A single particle momentum spectrum in the GCE
reads:
\eq{\label{Boltz}
F_{\rm gce}(p)~\equiv~\frac{1}{\overline{N}}
~\frac{dN}{p^2dp}~=~\frac{V}{2\pi^2~\overline{N}}~\exp\left(-~\frac{p}{T}\right)
~=~\frac{1}{2T^3}~\exp\left(-~\frac{p}{T}\right)~.
}
The MCE spectrum is very close to the Boltzmann
distribution (\ref{Boltz})
at all  momenta up to 10~GeV/c. At this momentum, the spectrum $F(p)$
is already dropped in a comparison to its $F(0)$ value by about a factor of $10^{-30}$.
The MCE spectrum decreases faster than the GCE one at high
momenta. Close to the threshold momentum, $p=E=24$~GeV, where the MCE
spectrum goes to zero, large deviations from (\ref{Boltz}) are
observed. In order to demonstrate these deviations the MCE and GCE momentum
spectra are shown in Fig.~\ref{fig-p-distr}  (b) over 90 orders of
magnitude.

\subsection{Hadron Resonance Gas in GCE, CE, and MCE}
\begin{figure}
\includegraphics[width=.45\textwidth]{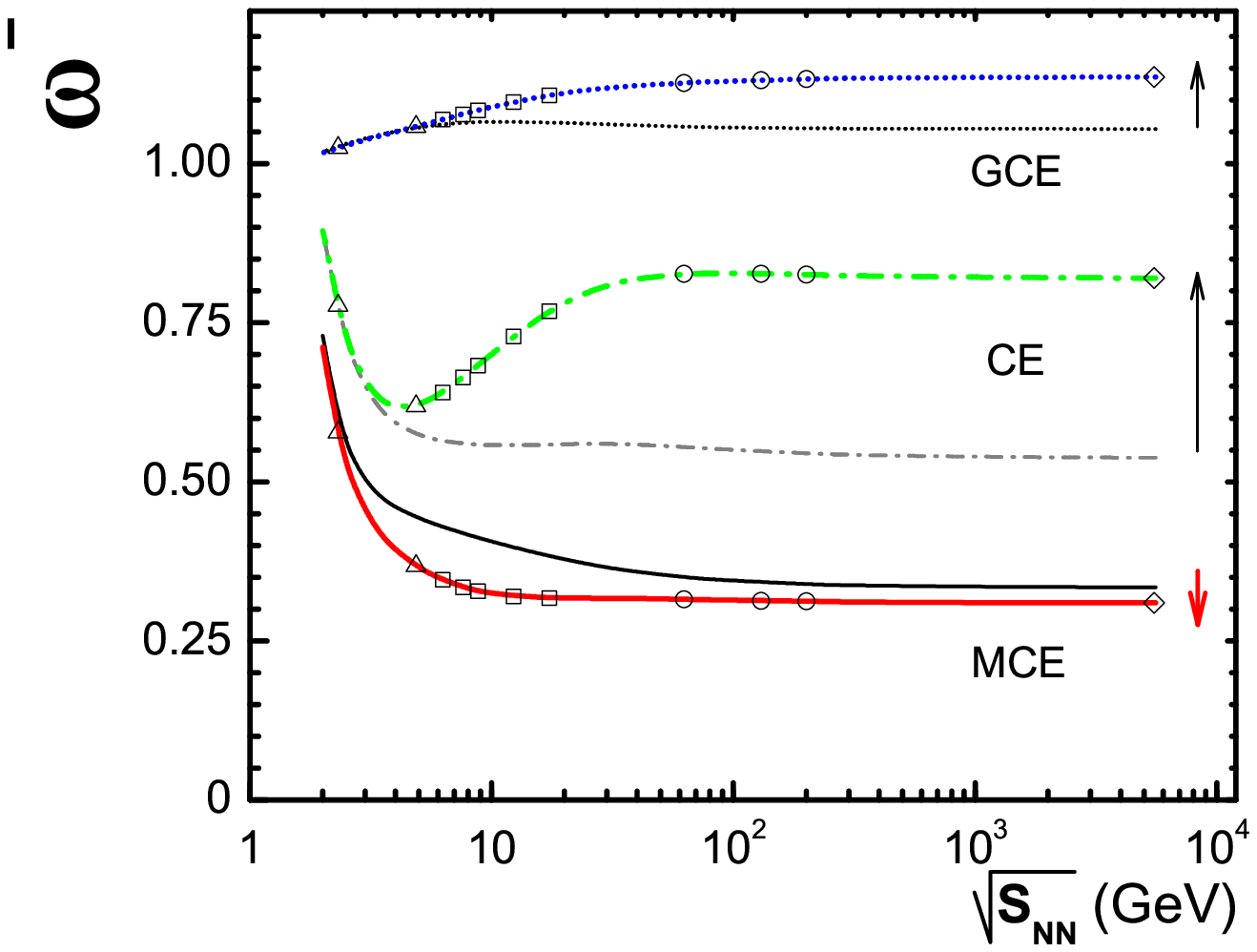}
\includegraphics[width=.45\textwidth]{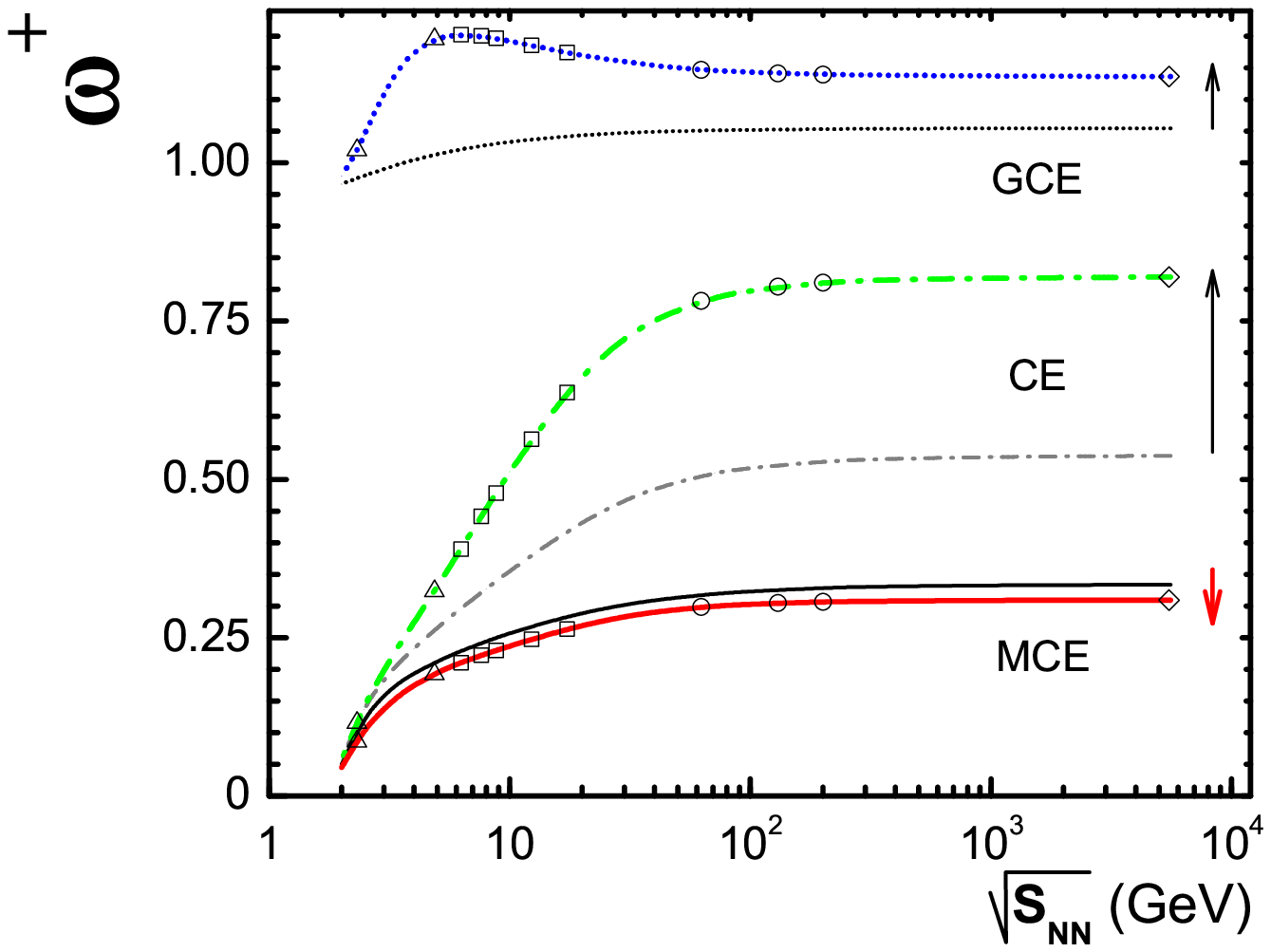}
\caption{The scaled variances $\omega^-$ ({\it left}) and $\omega^+$ ({\it right}),
both primordial and final, along the chemical freeze-out line for
central Pb+Pb (Au+Au) collisions. Different lines present the GCE, CE, and MCE
results. Symbols at the curves for final particles correspond to the specific collision
energies. The arrows show the effect of resonance decays.
 } \label{fig-omega-pm}
\end{figure}

We present now the results of the hadron resonance gas (HRG) for
the e-by-e fluctuations of negatively charged
and positively charged hadrons (see details in Ref.~\cite{exp}).
The corresponding scaled variances
$\omega^-$ and $\omega^+$ are calculated in the GCE, CE, and MCE
along the chemical freeze-out line in central Pb+Pb (Au+Au) collisions
for the whole energy range from SIS to LHC. The model parameters
are the volume $V$, temperature $T$, baryonic chemical potential $\mu_B$,
and the strangeness saturation parameter $\gamma_S$. They are chosen by
fitting the mean hadron multiplicities.
Once a suitable set of the chemical freeze-out parameters is determined
for each collision energy, the scaled variances $\omega^-$ and $\omega^-$
can be calculated in different statistical ensembles. The results
for the GCE, CE, and MCE are presented in Fig.~\ref{fig-omega-pm}
as functions of the center-of-mass energy $\sqrt{s_{NN}}$ of the nucleon pair.
\begin{figure}
\includegraphics[width=.45\textwidth]{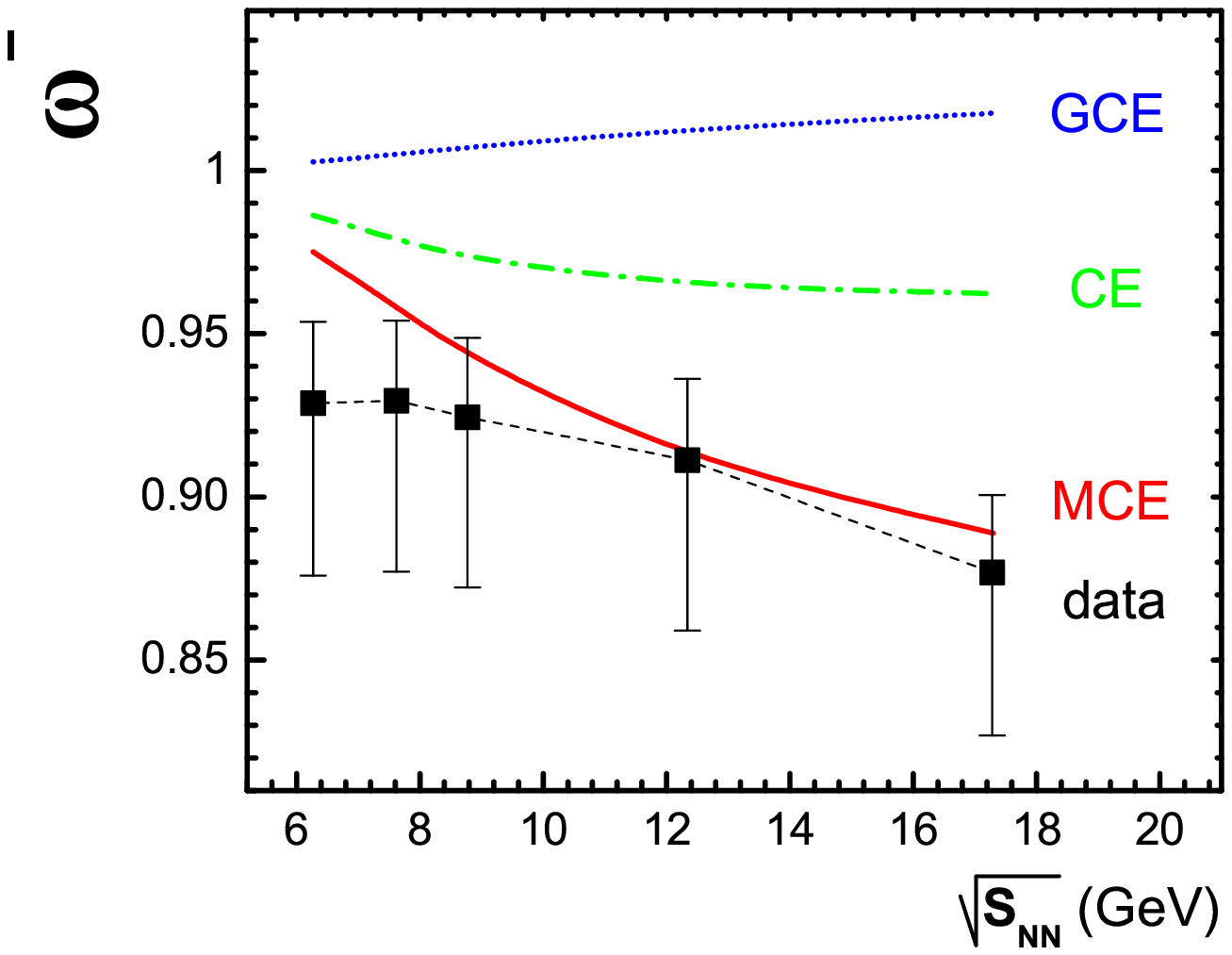}
\includegraphics[width=.45\textwidth]{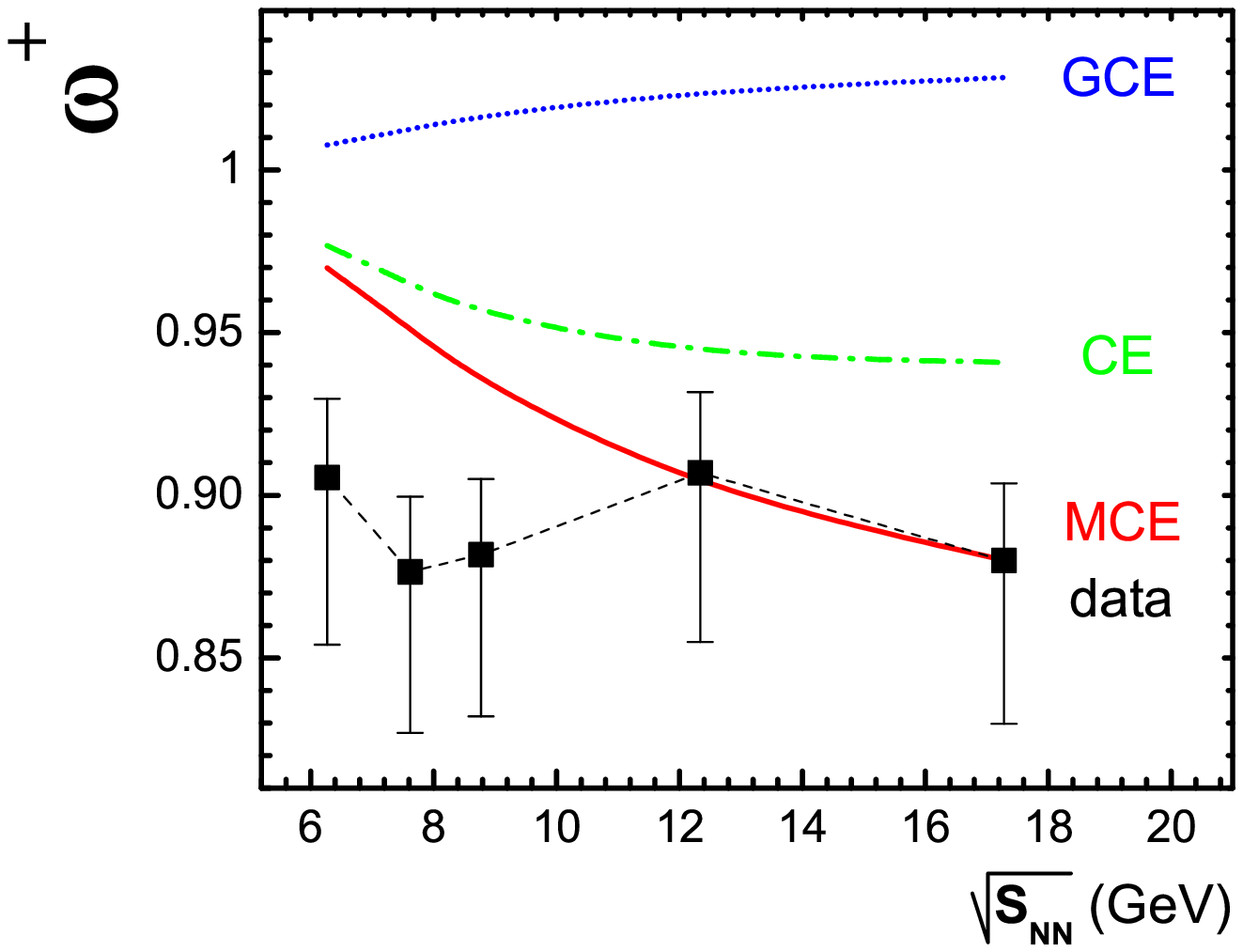}
\caption{The scaled variance for negatively charged ({\it left}) and
positively charged ({\it right}) hadrons along the chemical freeze-out
line for central Pb+Pb collisions at the SPS energies. The points show the data
of the NA49 Collaboration \cite{NA49-Lung}. Curves show the GCE, CE, and MCE
results calculated with the NA49 acceptance.
 } \label{fig-omega-pm-data}
\end{figure}
A comparison between the data and predictions of statistical models
should be performed for results which correspond to A+A collisions
with a fixed number of nucleon participants. In Fig.~\ref{fig-omega-pm-data}
our statistical model results are compared with the NA49 data \cite{NA49-Lung}
at collision energies 20, 30, 40, 80 and 158~A~GeV
for the 1\% most central Pb+Pb collisions selected by
the numbers of projectile participants.
In the experimental study of A+A collisions at high energies only
a fraction of all produced particles is registered. If detected
particles are uncorrelated, the scaled variance for the accepted particles,
$\omega_{\rm acc}[N]$, can be presented in terms of the full 4$\pi$
scaled variance $\omega[N]$
as (see, e.g., \cite{exp})
\eq{\label{acc}
\omega_{\rm acc}[N]~=~1~-~q~+~q\,\omega[N]~,
}
where $q$ is a probability of a single particle to be accepted.
Figure \ref{fig-omega-pm-data} demonstrates that among three different
statistical ensembles the MCE
is in a better agreement with the NA49 data. The reasons of this
are not yet clear.

\subsection{Generalized Statistical Ensembles}\label{ss-gse}
A choice of statistical ensemble is crucial in calculating
fluctuations. On the other hand, it is clear that GCE, CE, and MCE
are only some typical examples.
A general concept of the  statistical ensembles was
suggested in Ref.~\cite{alpha}. The
extensive quantities $(V,E,Q)\equiv\vec{A}$ define the MCE.
Different statistical ensembles are then constructed
using externally given distributions of extensive
quantities, $P_{\alpha}(\vec{A})$.  The
distribution of any observable $O$ in is then obtained in
two steps. Firstly, the MCE $O$-distribution,
$P_{mce}(O;\vec{A})$,  is calculated at fixed values of the
extensive quantities $\vec{A}$. Secondly, this result is
averaged over the external distribution $P_{\alpha}(\vec{A})$ \cite{alpha}:
\begin{eqnarray}\label{P}
P_{\alpha} (O) ~=~ \int  d\vec{A}  ~P_{\alpha}(\vec{A}) ~
P_{mce}(O;\vec{A})~.
\end{eqnarray}
Fluctuations of extensive quantities $\vec{A}$ around
their average values depend not on the system's physical properties,
but rather on external conditions. One can imagine a huge
variety of these conditions, thus, the standard statistical ensembles
discussed above are only some special examples.
The ensemble defined by Eq.~(\ref{P}), the $\alpha$-ensemble,
includes the standard statistical ensembles as the particular cases.
The generalized statistical mechanics based on Eq.~(\ref{P}) can be applied
to different tasks of hadron production in high energy collisions.
For example, based on Eq.~(\ref{P}) and introducing the scaling volume
fluctuations $P_{\alpha}(V)$, an attempt was made in Ref.~\cite{MCEsVF}
to extend the statistical model
to the hard domain of high transverse momenta and/or high hadron masses.

\section{Strongly Intensive Measures of E-by-E Fluctuations}\label{s-SIQ}
A significant increase of transverse momentum and
multiplicity fluctuations is expected in the vicinity of the
CP. One can probe different regions of the
phase diagram by varying the collision energy and the size of
colliding nuclei. The possibility to observe signatures
of the critical point inspired the energy and system size scan
program of the NA61/SHINE Collaboration at the CERN
SPS \cite{Ga:2009} and the low beam energy scan program
of the STAR and PHENIX Collaborations at the
BNL RHIC \cite{RHIC-SCAN}. In these studies one measures and then compares
e-by-e
fluctuations in collisions of different nuclei at
different collision energies. The average sizes of the created
physical systems and their e-by-e fluctuations are expected to be
rather different (see, e.g., Ref.~\cite{KGGB}). This strongly affects the
observed fluctuations, i.e., the measured quantities would
not describe the local physical properties of the system but
rather reflect the system size fluctuations. For instance, A+A
collisions with different centralities may produce a system with
approximately the same local properties (e.g., the same
temperature and baryonic chemical potential) but with the volume
changing significantly from interaction to interaction. Note that
in high energy collisions the average volume  of created matter
and its variations from collision to collision usually cannot be
controlled experimentally.
Therefore, a suitable choice of
statistical tools for the study of e-by-e fluctuations is really
important.

Intensive quantities are defined within the GCE
of statistical mechanics. They depend on temperature and
chemical potential(s), but they are independent of the system
volume.
Strongly intensive quantities introduced in Ref.~\cite{GG:2011} are, in addition,
independent of volume fluctuations.
They are the appropriate
measures for studies of e-by-e
fluctuations in A+A collisions
and can be defined for two
extensive state quantities $A$ and $B$. Here, we call $A$
and $B$ {\it extensive} when the first moments of their
distributions for the ensemble of possible states are proportional
to volume. They are referred to as {\it state quantities} as they
characterize the states of the considered system, e.g., final
states
of A+A collisions or
micro-states of the GCE.

There are two families of strongly intensive quantities which
depend on the second and first moments of $A$ and $B$ and thus
allow to study e-by-e (or state-by-state) fluctuations~\cite{GG:2011}:
 \eq{\label{Delta-AB}
 \Delta[A,B]
 ~&=~ \frac{1}{C_{\Delta}} \Big[ \langle B\rangle\,
      \omega[A] ~-~\langle A\rangle\, \omega[B] \Big]~,
 \\
  \Sigma[A,B]
 ~&=~ \frac{1}{C_{\Sigma}}\Big[
      \langle B\rangle\,\omega[A] ~+~\langle A\rangle\, \omega[B] ~-~2\left(
      \langle AB \rangle -\langle A\rangle\langle
      B\rangle\right)\Big]~,
 \label{Sigma-AB}
 }
where
 \eq{\label{omega-AB}
 \omega[A] ~\equiv~ \frac{\langle A^2\rangle ~-~ \langle A\rangle^2}{\langle
 A\rangle}~,~~~~
 \omega[B] ~\equiv~ \frac{\langle B^2\rangle ~-~ \langle B\rangle^2}{\langle
 B\rangle}~,
 }
and averaging $\langle \ldots \rangle$ is performed over the
ensemble of multi-particle states.
The normalization factors
$C_{\Delta}$ and $C_{\Sigma}$ are required to be proportional to
the first moments of any extensive quantities.

In Ref.~\cite{GGP:2013} a specific choice of the $C_{\Delta}$ and
$C_{\Sigma}$ normalization factors was proposed. It  makes the quantities
$\Delta[A,B]$ and $\Sigma[A,B]$ dimensionless and  leads to
$\Delta[A,B] = \Sigma[A,B] = 1$ in the independent particle
model,  as will be shown below.

From the definition of $\Delta[A,B]$ and $\Sigma[A,B]$ it follows that
$\Delta[A,B] = \Sigma[A,B] = 0$ in the case of absence of
fluctuations of $A$ and $B$, i.e., for $ \omega[A] = \omega[B] =
\langle AB \rangle -\langle A\rangle \langle B \rangle = 0 $. Thus
the proposed normalization of $\Delta[A,B]$ and $\Sigma[A,B]$
leads to a common scale on which the values of the fluctuation
measures calculated for different state quantities $A$ and $B$ can
be compared.

There is an important  difference between the $\Sigma[A,B]$ and
$\Delta[A,B]$ quantities. Namely, in order to calculate
$\Delta[A,B]$ one  needs to measure only the first two moments:
$\langle A\rangle$, $\langle B\rangle$ and $\langle A^2\rangle$,
$\langle B^2\rangle$. This can be done by  independent
measurements of the distributions $P_A(A)$ and $P_B(B)$. The
quantity $\Sigma[A,B]$ includes the correlation term, $\langle AB
\rangle -\langle A \rangle \langle B \rangle$, and thus
requires, in addition,  simultaneous measurements of $A$ and $B$
in order to obtain the joint distribution $P_{AB}(A,B)$.

\subsection{$\Delta$ and $\Sigma$ in the Independent Particle Model }\label{ss-IPM}

The IPM assumes that:

(1) the state
quantities $A$ and $B$ can be
expressed as
\eq{\label{ind-part}
A~=~\alpha_1~+\alpha_2~+\ldots~+
\alpha_{N}~,~~~~~B~=~\beta_1~+\beta_2~+\ldots~+ \beta_{N}~,
}
where  $\alpha_j$ and $\beta_j$ denote single particle
contributions to $A$ and $B$, respectively, and $N$ is the number
of particles;

(2) inter-particle correlations are absent, i.e., the probability
of any multi-particle state is the product of probability
distributions $P(\alpha_j,\beta_j)$ of single-particle states, and
these probability distributions are the same for all $j=1,\ldots,
N$ and independent of $N$,
\eq{ \label{dist-alpha} P_N (\alpha_1,\beta_1,\alpha_2,\beta_2,
\dots, \alpha_N,\beta_N) = {\cal P}(N)\times
P(\alpha_1,\beta_1)\times  P(\alpha_2,\beta_2)\times \cdots \times
P(\alpha_N,\beta_N)~, }
where ${\cal P}(N)$ is an arbitrary multiplicity distribution of
particles.

It can be shown \cite{GGP:2013} that within the
IPM the average values of the first and second moments of $A$ and
$B$ are equal to:
 \eq{\label{A}
 &\langle A \rangle~=~\overline{\alpha} ~\langle N\rangle~,~~~~
 \langle A^2 \rangle~=~\overline{\alpha^2}~\langle N \rangle ~+~
 \overline{\alpha}^2
 ~\left[\langle N^2\rangle ~-~\langle N\rangle\right]~,\\
&\langle B \rangle~=~\overline{\beta}~\langle N\rangle~,~~~~
 \langle B^2 \rangle~=~\overline{\beta^2}~\langle N \rangle ~+~
 \overline{\beta}^2
 ~\left[\langle N^2\rangle ~-~\langle N\rangle\right]~,\label{B} \\
 &\langle AB \rangle~=~
\overline{\alpha\,\beta}~\langle N \rangle ~+~
 \overline{\alpha}\,\cdot\,\overline{\beta}
 ~\left[\langle N^2\rangle ~-~\langle
 N\rangle\right]~.\label{AB}
}
The values of $\langle A \rangle$ and $\langle B \rangle$ are
proportional to the average number of particles $\langle N
\rangle$ and, thus, to the average size  of the system. These
quantities are extensive.  The quantities $\overline{\alpha}$,
~$\overline{\beta}$ and $\overline{\alpha^2}$,
~$\overline{\beta^2}$, ~$\overline{\alpha\,\beta}$ are the first
and second moments of the single-particle distribution
$P(\alpha,\beta)$. Within the IPM they are independent of $\langle
N \rangle$ and play the role of intensive quantities.

Using Eq.~(\ref{A}) the scaled variance $\omega[A]$ which
describes the state-by-state fluctuations of $A$ can be expressed
as:
 \eq{\label{omega-A}
  \omega[A]~\equiv~ \frac{\langle A^2\rangle
~-~\langle A \rangle ^2}{\langle
A\rangle}~=~\frac{\overline{\alpha^2}~-~\overline{\alpha}^2}
{\overline{\alpha}}~+~\overline{\alpha}~\frac{\langle N^2\rangle
~-~\langle N\rangle^2}{\langle N\rangle}
~\equiv~\omega[\alpha]~+~\overline{\alpha}~ \omega[N]~,
}
where $\omega[\alpha]$ is the scaled variance of the
single-particle quantity $\alpha$, and $\omega[N]$ is the scaled
variance of $N$.
A similar expression follows from Eq.~(\ref{B}) for the scaled
variance $\omega[B]$. The scaled variances $\omega[A]$ and
$\omega[B]$ depend on the fluctuations of the particle number via
$\omega[N]$. Therefore, $\omega[A]$ and $\omega[B]$ are not
strongly intensive quantities.

From Eqs.~(\ref{A}-\ref{AB}) one obtains expressions for
$\Delta[A,B]$ and $\Sigma[A,B]$, namely:
\eq{
\Delta[A,B]~&=~ \frac{\langle N \rangle}{C_{\Delta}}~
\Big[~\overline{\beta}~ \omega[\alpha] ~-~\overline{\alpha}
~\omega[\beta]~\Big]~, \label{IPM-D}\\
\Sigma[A,B]~&=\frac{\langle N
\rangle}{C_{\Sigma}}~\Big[~\overline{\beta}~\omega[\alpha]~+~\overline{\alpha}~
\omega[\beta] ~-~2\left(~\overline{\alpha\, \beta}
-\overline{\alpha}\cdot \overline{\beta}~\right)~\Big]~.
\label{IPM-S}
}
Thus, the requirement that
\eq{\label{DS=1}
\Delta[A,B]~ = ~\Sigma[A,B]~ =~ 1~,
}
within the IPM leads to:
\eq{
C_{\Delta}~&=~\langle N \rangle~ \Big[~\overline{\beta}~
\omega[\alpha] ~-~\overline{\alpha}
~\omega[\beta]~\Big]~, \label{C-D}\\
C_{\Sigma}~&=~\langle N
\rangle~\Big[~\overline{\beta}~\omega[\alpha]~+~\overline{\alpha}~
\omega[\beta] ~-~2\left(~\overline{\alpha\, \beta}
-\overline{\alpha}\cdot \overline{\beta}~\right)~\Big]~.
\label{C-S}
}

In the IPM the  $A$ and $B$ quantities are expressed in terms of
sums of the single particle variables, $\alpha$ and $\beta$. Thus
in order to calculate the normalization $C_{\Delta}$ and
$C_{\Sigma}$ factors one has to measure the single particle
quantities $\alpha$ and $\beta$. However, this may not  always be
possible within a given experimental set-up. For example, $A$ and
$B$ may be energies of particles measured by two calorimeters.
Then one can study fluctuations in terms of $\Delta[A,B]$ and
$\Sigma[A,B]$ but can not calculate the normalization factors which
are proposed above.

We consider now two examples of
specific pairs of extensive
variables $A$ and $B$. In the first example, we use the transverse momentum $P_T=p_t^{(1)}+\dots
p_t^{(N)}$, where $p^{(i)}_t$ is the absolute value of the
$i^{{\rm th}}$ particle transverse momentum, and the number of
particles $N$.
The requirement that
\eq{\Delta[P_T,N]~=~\Sigma[P_T,N]~=~1 \label{DelSig=1}
}
for the IPM leads then to the normalization factors \cite{GGP:2013}
\eq{\label{norm}
 C_{\Delta}~=~C_{\Sigma}=~
\omega[p_t]\cdot \langle N\rangle~,~~~~~
 \omega[p_t]~\equiv~\frac{\overline{p_t^2}~-~\overline{p_t}^2}
{\overline{p_t}}~,
}
where $\omega[p_t]$ describes the single particle $p_t$-fluctuations.

As the second example let us consider the multiplicity fluctuations. Here $A$ and $B$
will denote the multiplicities of hadrons of types $A$ and $B$, respectively (e.g.,
kaons and pions).
One obtains \cite{GGP:2013}
%
\eq{\label{CAB}
C_{\Delta}=\langle B\rangle -\langle A\rangle~,~~~~C_{\Sigma}=\langle A\rangle+\langle B\rangle~.
}
The normalization factors (\ref{norm}) and (\ref{CAB}) are suggested to be used for the calculation
for $\Delta$ and $\Sigma$ both in theoretical models and for the analysis of
experimental data (see Ref.~\cite{GGP:2013} for further details of the normalization
procedure).

The $\Phi$ measure, introduced some time ago~\cite{GM:1992},
belongs to the $\Sigma$ family within the current classification scheme. The fluctuation measure $\Phi$
was introduced for the study of transverse momentum fluctuations.
In the general case, when $A=X$  represents
any motional variable and $B=N$ is the particle multiplicity, one gets:
\eq{\label{Phi}
\Phi_X~=~\Big[\overline{x}\,\omega[x]\Big]^{1/2}\,\Big[\sqrt{\Sigma[X,N]}
~-~1\Big]~.
}
For the multiplicity fluctuations of hadrons belonging to non-overlapping types $A$ and $B$
the connection between the $\Phi[A,B]$ and $\Sigma[A,B]$ measures reads:
\eq{\label{phiAB}
\Phi[A,B]~=~\frac{\sqrt{\langle A\rangle\,\langle B\rangle}}{\langle A\rangle +\langle B\rangle}\,
\Big[\sqrt{\Sigma[A,B]}~-~1\,\Big]~.
}

The IPM plays an important role as the {\it reference
model}. The deviations of real data from the IPM
results Eq.~(\ref{DelSig=1}) can be used to learn about the physical
properties of the system. This resembles the situation in
studies of particle multiplicity fluctuations.
In this case, one uses the Poisson distribution
$P(N)=\exp\left(-\,\overline{N}\right)\,\overline{N}^N/{N!}$~
with $\omega[N]=1$ as the reference model. The other reference value
$\omega[N]=0$ corresponds to $N={\rm const}$, i.e., the absence
of $N$-fluctuations.
Values of $\omega[N]>1$ (or $\omega[N]\gg 1$) correspond to ``large''
(or ``very large'') fluctuations of $N$, and $\omega[N]<1$
(or $\omega[N]\ll 1$) to ``small''
(or ``very small'') fluctuations.

The fluctuation measures $\Delta$ and $\Sigma$ do not depend on
the average size of the system and its fluctuations
in several different model approaches, namely, statistical mechanics
within the GCE, model of Multiple Independent Sources, Mixed Event Model
(see Ref.~\cite{GGP:2013} for details). Thus, one may expect that
$\Delta$ and $\Sigma$  preserve their strongly intensive properties in
many real experiments too. An extension of the strongly intensive
fluctuation measures for higher order cumulants were suggested in recent paper
\cite{cumul}.

\subsection{$\Delta$ and $\Sigma$ Evaluated in Specific Models}\label{ss-SIQ_in_models}
The UrQMD  \cite{urqmd} calculations of $\Delta$ and $\Sigma$ measures were performed
in Refs.~\cite{KG:2012,Be:2012}. The Monte Carlo simulations and analytical model
results for $\Delta[P_T,N]$ and $\Sigma[P_T,N]$ were presented in Ref.
\cite{GorGr:2013}.
These
measures
were also studied in
Ref.~\cite{GR:2013} for the ideal Bose and Fermi gases within the
GCE. The GCE for the Boltzmann
approximation satisfies the conditions of the IPM, i.e.,
Eq.~(\ref{DS=1}) is valid. The following general
relations were found \cite{GR:2013}:
\eq{\label{BBF}
& \Delta^{{\rm Bose}}[P_T,N]~<~\Delta^{{\rm Boltz}}=1~<~\Delta^{{\rm Fermi}}[P_T,N]~,\\
& \Sigma^{{\rm Fermi}}[P_T,N]~<~\Sigma^{{\rm
Boltz}}=1~<~\Sigma^{{\rm Bose}}[P_T,N]~,\label{FBB}
}
i.e.,  the Bose statistics makes $\Delta[P_T,N]$ smaller and
$\Sigma[P_T,N]$ larger than unity, whereas the Fermi statistics
works in the opposite way.
The Bose statistics of pions appears to be the main source of
quantum statistics effects in a hadron gas with a temperature
typical for the hadron system created in A+A collisions. It gives
about 20\% decrease of $\Delta[P_T,N]$ and 10\% increase of
$\Sigma[P_T,N]$, at $T\cong 150$~MeV with respect to the IPM
results (\ref{DS=1}). The Fermi statistics of protons modifies
insignificantly $\Delta[P_T,N]$ and $\Sigma[P_T,N]$ for
typical values of $T$ and $\mu_B$. Note that UrQMD takes into account
several sources of fluctuations and correlations, e.g., the exact
conservation laws and resonance decays. On the other hand, it does
not include the effects of Bose and Fermi statistics.
First experimental results on $\Delta[P_T,N]$ and $\Sigma[P_T,N]$
in p+p and Pb+Pb collisions have been reported in Refs.~\cite{NA49_Phipt,ismd2013,NA61}.

Now we consider the $\Delta$ and $\Sigma$ measures for two particle multiplicities
$N_1$ and $N_2$. Resonance decays, when particle species 1 and 2 appear simultaneously
among the decay products, lead to the (positive) correlations between $N_1$ and $N_2$
numbers.
Let $N_1=\pi^+$ and $N_2=\pi^-$ are the  multiplicities
of positively and negatively charged  pions, respectively.
A presence of two components
is assumed: the correlated pion pairs coming from decays,
$R_{\pi\pi}\rightarrow \pi^++\pi^-$,
and the uncorrelated $\pi^+$ and $\pi^-$
from other sources. The $\pi^+$ and $\pi^-$ numbers are then equal to:
\eq{
\pi^+~=~n_+~+~R_{\pi\pi}~,~~~~~~\pi^-~=~n_-~+~R_{\pi\pi}~,
}
where $R_{\pi\pi}$ is the number of resonances decaying
into $\pi^+\pi^-$ pairs, while $n_+$ and $n_-$ are the numbers
of uncorrelated
$\pi^+$ and $\pi^-$, respectively.

Using approximate relations,
\eq{\label{omegas-pm}
%
\omega[\pi^+]~\cong ~\omega[\pi^-]~
\cong ~\omega[R_{\pi\pi}] \cong 1~,~
}
one obtains
two alternative expressions for the number of $R_{\pi\pi}$-resonances \cite{BGG}:
\eq{
\label{R-1}
& \frac{\langle R_{\pi\pi} \rangle}{\langle \pi^-\rangle +\langle \pi^+ \rangle}~\cong~
\frac{\langle \pi^+\pi^-\rangle~-~\langle \pi^+\rangle \langle \pi^-\rangle}
{\langle \pi^+\rangle~+~\langle{\pi^-}\rangle}~\equiv ~\rho[\pi^+,\pi^-]~,\\
&\frac{\langle R_{\pi\pi} \rangle}{\langle \pi^-\rangle +\langle \pi^+ \rangle}~\cong~
\frac{1 ~-~ \Sigma[\pi^+,\pi^-]}{2}~,\label{R-2}
%
}
i.e., $R_{\pi\pi}$ can be calculated using the
measurable quantities of e-by-e fluctuations.
As a result, these resonance abundances, which are difficult to be measured by other methods,
can be estimated by measuring the fluctuations and correlations of the numbers of stable hadrons.
Note that an idea to use the e-by-e fluctuations
of particle number ratios to estimate the number of hadronic resonances was
suggested for the first time
in Ref.~\cite{JK}.

The analysis of $\pi^+$ and $\pi^-$ fluctuations and correlations from resonance decays is
done in Ref.~\cite{BGG}. It is based on the HRG model, both
in the GCE and CE, and on the relativistic transport model UrQMD.
This analysis illustrates a role of the centrality selection,
limited acceptance, and global charge conservation in A+A collisions.
We present now the  results for Eq.~(\ref{R-1}) and (\ref{R-2}) in Pb+Pb and p+p collisions.
Note that $\rho$ in Eq.~(\ref{R-1}) is an intensive but not strongly intensive
quantity. Thus, it is expected to be sensitive to the system size fluctuations.
On the other hand, the $\Sigma$ in Eq.~(\ref{R-2}) is the strongly intensive measure
and it should keep the same value in a presence of the system size fluctuations.

The samples of
5\% central Pb+Pb collision events at $\sqrt{s_{NN}}= 6.27$ and 17.3~GeV are considered.
These UrQMD results are  compared with those for
most central Pb+Pb collisions at zero impact parameter, $b=0$~fm,
and for  $p+p$ reactions at the same collision energies.
Several mid-rapidity windows
$-\Delta y/2 <y<\Delta y/2$ for final  $\pi^+$ and $\pi^-$ particles
are considered.

A width of the rapidity window $\Delta y$ is an important parameter.
The two hadrons which are the products of a resonance decay have, in average, a
rapidity difference of the order of unity. Therefore, while searching for the effects of resonance
decays one should choose $\Delta y\geq 1$
to enlarge a probability for simultaneous hit into the rapidity window $\Delta y$
of both  correlated hadrons  (e.g., $\pi^+$ and $\pi^-$) from resonance decays.
Thus, $\Delta y$ should be {\it large} enough.  However, $\Delta y$ should be
{\it small} in  comparison to the whole rapidity interval $\Delta Y \approx \ln(\sqrt{s_{NN}}/m)$
accessible for final hadron with mass $m$.
 Only for  $\Delta y\ll \Delta Y$ one can expect a validity of
the GCE results.
Considering a small part of the statistical system, one does not need to impose
the restrictions of the exact global charge conservations: the GCE which only regulates
the average values of the conserved charges  is fully acceptable.
For large $\Delta y$, when the detected hadrons correspond to an essential
part of the whole system, the effects of the global charge conservation become more important.
In the HRG this should be treated within the CE, where the conserved charges
are fixed for all microscopic states. The global charge conservation influences
the particle number fluctuations  and introduces additional
correlations between numbers of different particle species.

\begin{figure}[t]
\centering
\includegraphics[width=0.495\textwidth]{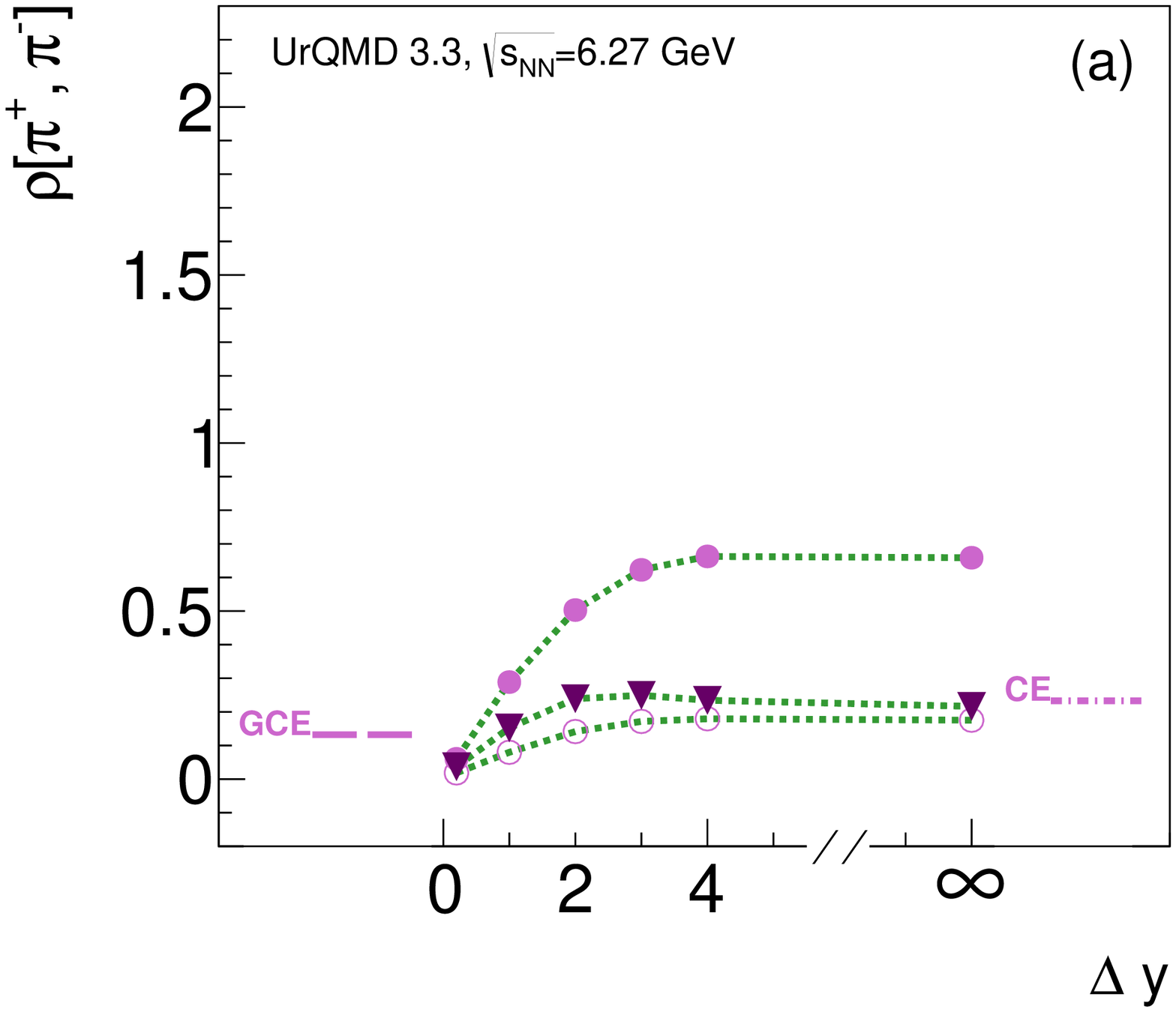}
\includegraphics[width=0.495\textwidth]{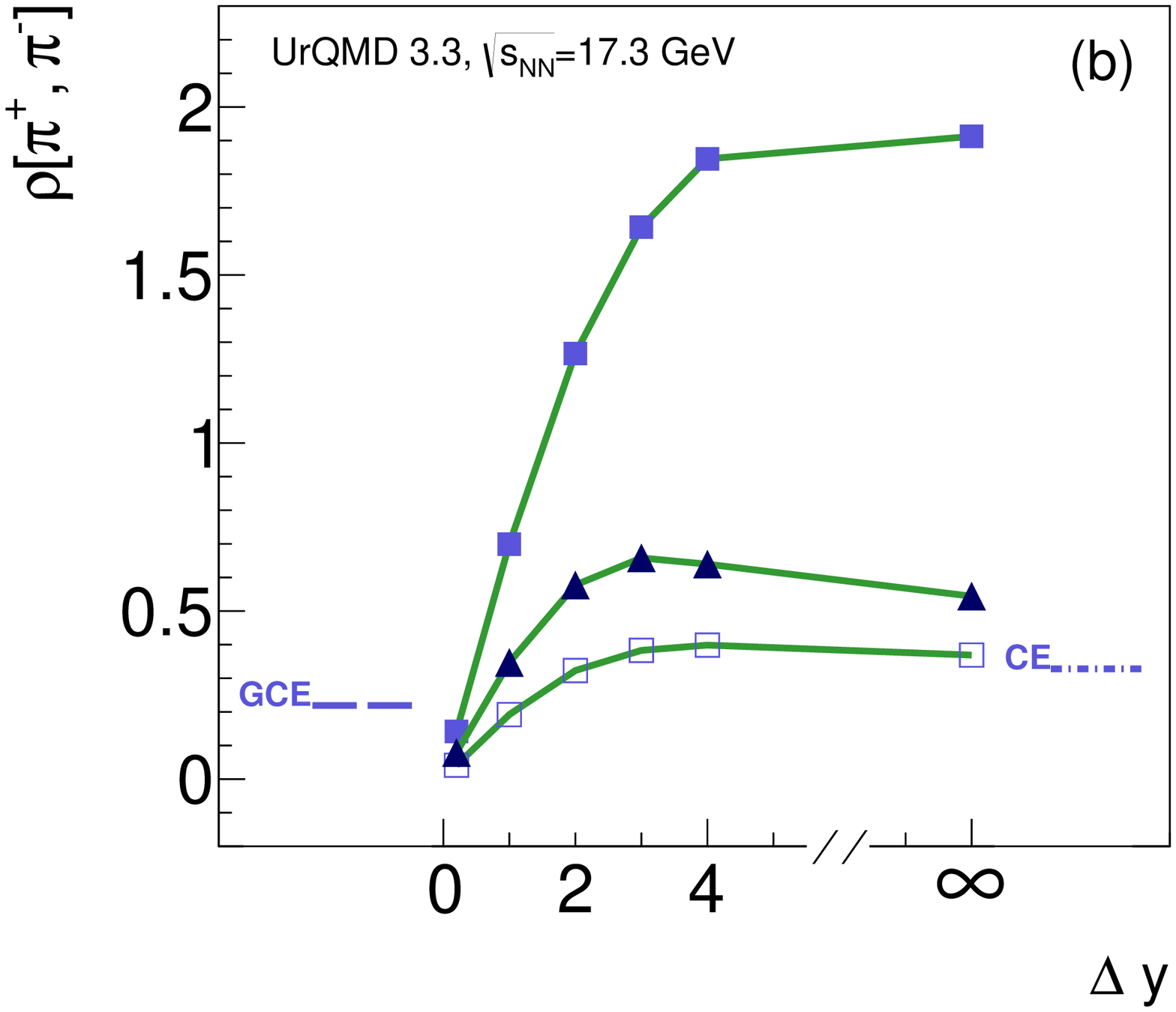}
\caption{ The UrQMD results for $\rho[\pi^+,\pi^-]$
in central Pb+Pb and inelastic $p+p$ collisions for the mid-rapidity windows
$\Delta y$, i.e., the center of mass rapidities $y$ of final $\pi^+$ and $\pi^-$
satisfy the condition $-\Delta Y< y< \Delta Y$.
Full circles and squares correspond to the 5\% centrality selection and  the open ones
to the most central Pb+Pb collision events with zero impact parameter $b=0$~fm.
Triangles correspond to the UrQMD simulations of inelastic $p+p$ interactions.
The collision energies are: $\sqrt{s_{NN}}=6.27$~GeV in ({\it a}) and  $\sqrt{s_{NN}}=17.3$~GeV
in ({\it b}). The horizontal
dashed and dashed-dotted lines show, respectively, the GCE and CE results
taken at the corresponding $\sqrt{s_{NN}}$.
}\label{fig-rho}
\end{figure}

\begin{figure}[t]
\centering
\includegraphics[width=0.495\textwidth]{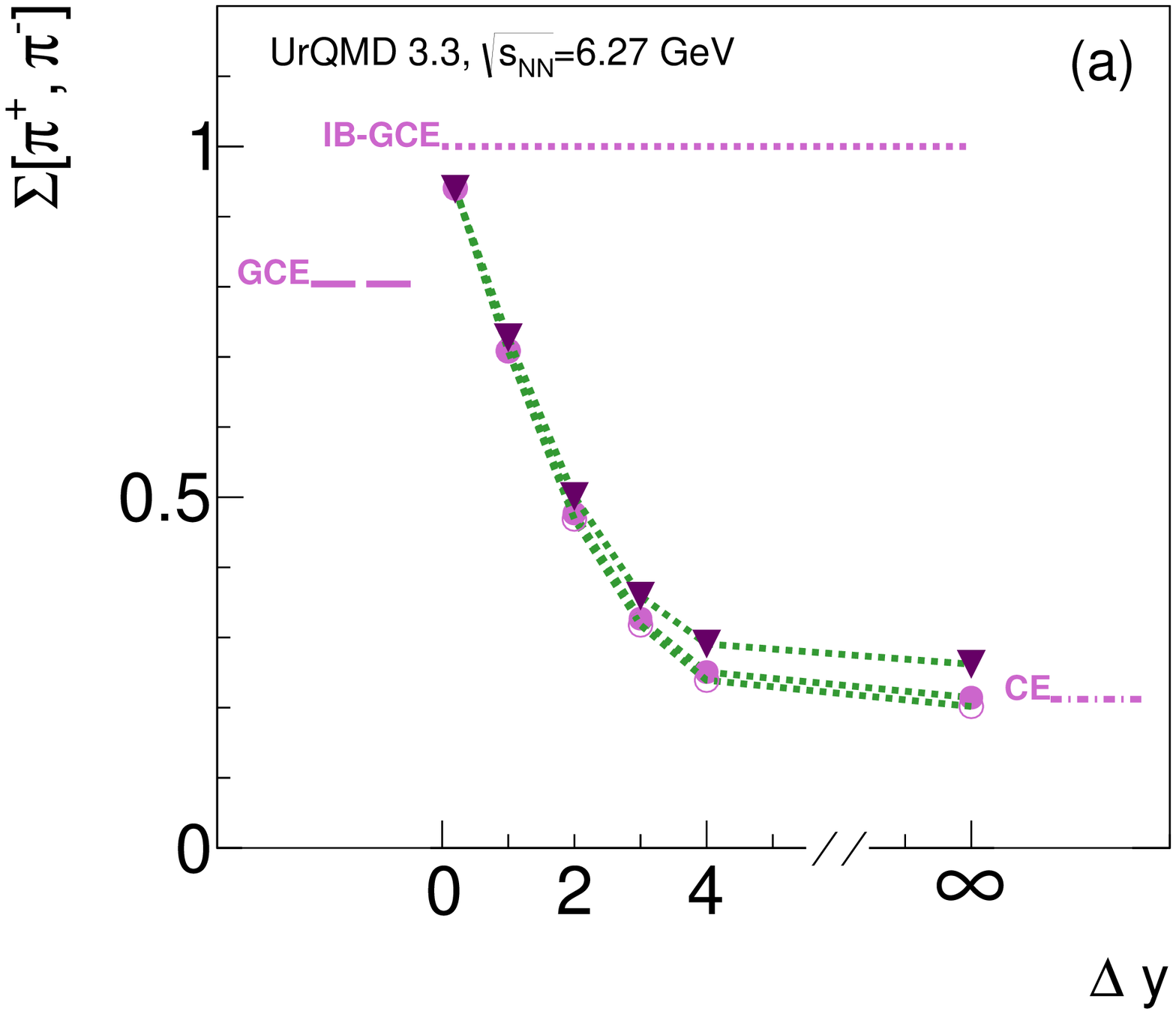}
\includegraphics[width=0.495\textwidth]{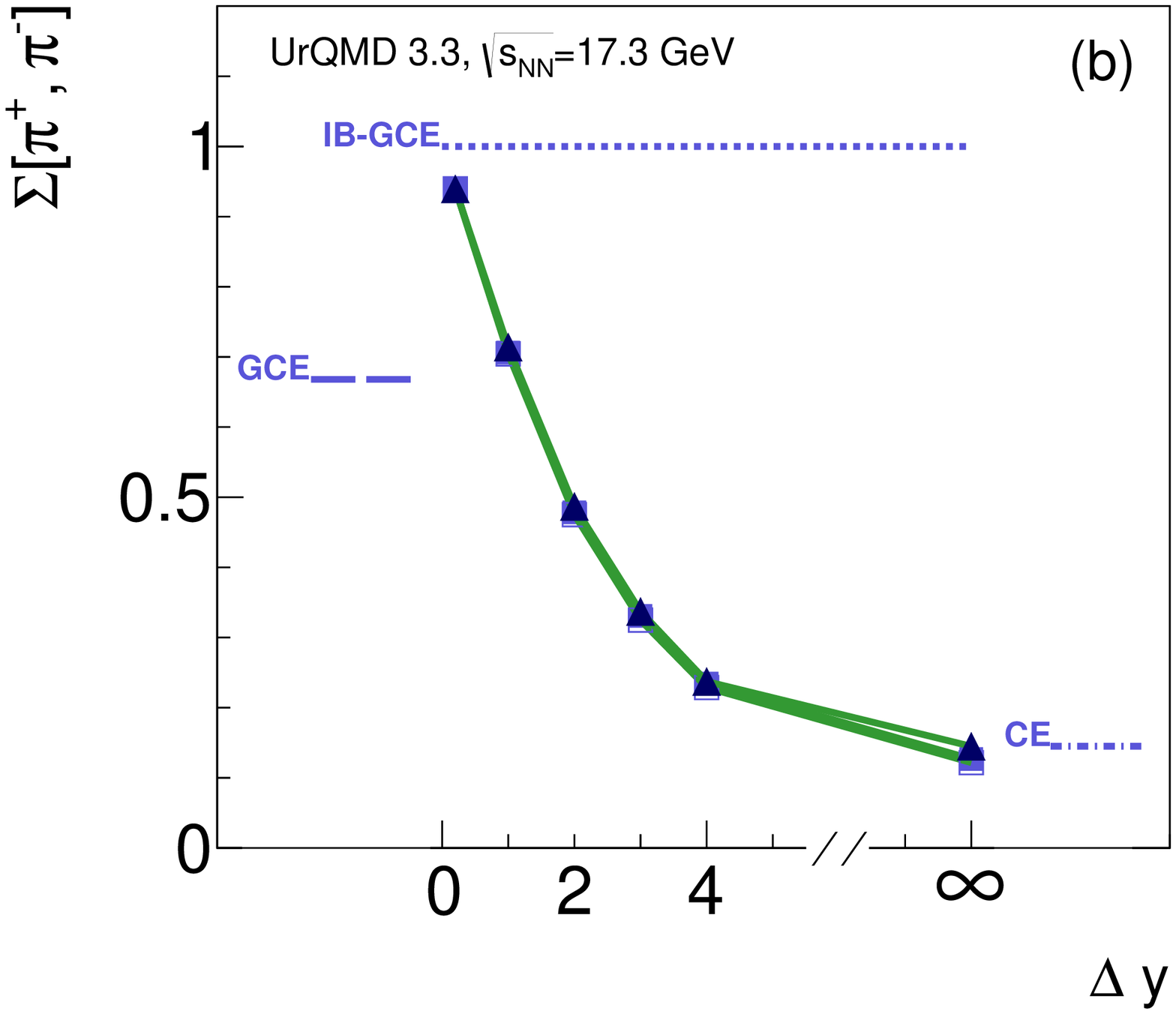}
\caption{ 
The
same as in Fig.~\protect\ref{fig-rho} but 
for $\Sigma[\pi^+,\pi^-]$.
}\label{fig-sigma}
\end{figure}
The UrQMD values of  correlation
parameter $\rho[\pi^+,\pi^-]$ in 5\% central Pb+Pb collision
events are shown by full circles in Fig.~\ref{fig-rho}(a) and full boxes
in \ref{fig-rho}(b)
as  functions of the acceptance
windows $\Delta y$. The UrQMD results for  the most central Pb+Pb collision events
with zero impact parameter, $b=0$~fm, are shown by open symbols. The triangles show
the results of the UrQMD simulations in $p+p$ reactions. The
collision energy is taken as $\sqrt{s_{NN}}=6.27$~GeV in Fig.~\ref{fig-rho}(a)
and $\sqrt{s_{NN}}=17.3$~GeV in Fig.~\ref{fig-rho}(b).
The windows at the center of mass mid-rapidity are taken as
$\Delta y= 0.2$, 1, 2, 3, 4, and $\infty$.
A symbol $\infty$ denotes the case when all final state particles are detected
(i.e., a full $4\pi$-acceptance).
Note that the UrQMD model does not assume
any, even local, thermal and/or chemical equilibration.
Therefore, a connection between the  UrQMD and HRG results for particle number fluctuations
and correlations is a priori unknown.

For the 5\% most central Pb+Pb collisions
the correlation $\rho[\pi^+,\pi^-]$
increases with $\Delta y$.  As seen from Fig.~\ref{fig-rho}(b),
this increase is rather strong at high collision energy:
at large $\Delta y$,
the value of  $\rho[\pi^+,\pi^-]$ becomes
much larger than the HGM results in both the
CE and GCE. The behavior of $\omega[\pi^+]$ and $\omega[\pi^-]$
is rather similar
to that of $\rho[\pi^+,\pi^-]$.

Selecting within the UrQMD simulations the most central Pb+Pb collision
events with zero impact parameter $b=0$~fm, one finds essentially smaller
values of $\rho[\pi^+,\pi^-]$ (open symbols in Fig.~\ref{fig-rho}).
This means that in the 5\% centrality bin of Pb+Pb collision events
large fluctuations of the number of
nucleon participants (i.e., the volume fluctuations) are present.
These volume fluctuations  produce large additional contributions
to the scaled variances of pions and to the correlation parameter $\rho[\pi^+,\pi^-]$.
They become more and more important with increasing
collision energy. This is due to an increase  of
the number of pions per participating nucleon with increasing collision energy.
However, one hopes that these volume fluctuations will be canceled out to a large
extent when they are combined in the strongly intensive measures.
Note also that the UrQMD results for  $\rho[\pi^+,\pi^-]$
in inelastic $p+p$ collisions,
shown in Fig.~\ref{fig-rho} by triangles, are qualitatively similar to those
in Pb+Pb collisions at $b=0$~fm.

The UrQMD results for $\Sigma[\pi^+,\pi^-]$ in Pb+Pb collisions
at $\sqrt{s_{NN}}=6.27$~GeV and 17.3~GeV are presented in Fig.~\ref{fig-sigma}(a) and (b),
respectively, as a function of the acceptance window
$\Delta y$ at mid-rapidity.
In contrast to the results shown in Fig.~\ref{fig-rho},
both centrality selections in Pb+Pb collisions (5\% centrality bin and $b=0$~fm)
lead to very similar results for $\Sigma[\pi^+,\pi^-]$ shown in Fig.~\ref{fig-sigma}.
This means that the measure $\Sigma[\pi^+,\pi^-]$ has the strongly intensive properties,
at least in the UrQMD
simulations. The UrQMD results in p+p
reactions are close to those in  Pb+Pb ones.

The GCE and CE results of the HRG
are presented in Figs.~\ref{fig-rho} and \ref{fig-sigma}
by the horizontal dashed and dashed-dotted lines, respectively.
The UrQMD results for $\Sigma[\pi^+,\pi^-]$, presented in Fig.~\ref{fig-sigma},
demonstrate a strong dependence on
the size of rapidity window $\Delta y$.
At  $\Delta y=1$ these results
are close to those of the GCE HRM.
On the other hand,
with increasing $\Delta y$ the role of exact charge conservation becomes more and
more important.
From Fig.~\ref{fig-sigma},  one observes
that the UrQMD values  of $\Sigma[\pi^+,\pi^-]$ at large $\Delta y$
are close to the CE results.

As seen from Fig.~\ref{fig-rho},
a similar correspondence between the UrQMD results for $\rho[\pi^+,\pi^-]$
in Pb+Pb collisions at $b=0$~fm
and their GCE and CE values  is approximately valid. However,
this is not the case for  the 5\% most central Pb+Pb events. In that centrality
bin the volume fluctuations give the dominant contributions to
$\rho[\pi^+,\pi^-]$  for large $\Delta y$.

For very small acceptance, $\Delta y \ll 1$,
one expects an approximate validity of the Poisson distribution for
any type of the detected particles. Their scaled variances are then
close to unity, i.e., $\omega[\pi^-]\cong \omega[\pi^+] \cong 1$.
Particle number correlations,
due to both the resonance decays and the global charge conservation,
become negligible, i.e., $\rho[\pi^+,\pi^-]\ll 1$.
These expectations are indeed supported by the UrQMD results at $\Delta y=0.2$
presented in Fig.~\ref{fig-rho}.
Therefore, one expects $\Sigma[\pi^+,\pi^-]\rightarrow 1$ at $\Delta y\rightarrow 0$.
This expectation is also valid, as seen from the UrQMD results at $\Delta y=0.2$
presented in Fig.~\ref{fig-sigma}.

\section{Incomplete Particle Identification}\label{s-IPI}

Fluctuations of the chemical (particle-type) composition of
hadronic final states in A+A collisions are expected to be
sensitive to the phase transition between hadronic and partonic
matter. First experimental results on e-by-e chemical fluctuations
have already been published from the CERN SPS and BNL RHIC, and more
systematic measurements are in progress.

Studies of chemical fluctuations in general require to determine the
number of particles of different hadron species (e.g., pions,
kaons, and protons) e-by-e.
%
A serious experimental problem in such measurements
is incomplete particle identification, i.e.,
the impossibility to identify uniquely the type of each
detected particle.
The effect of particle misidentification distorts the measured
fluctuation quantities. For this reason the analysis of chemical
fluctuations is usually performed in a small acceptance, where
particle identification is relatively reliable. However, an
important part of the information on e-by-e fluctuations in  full
phase space is then lost.

Although it is usually impossible to identify each detected
particle, one can nevertheless determine with high accuracy the
average multiplicities (averaged over many events) for different
hadron species.

\subsection{The Identity Method}\label{ss-IM}
The identity variable was introduced in Ref.~\cite{Ga:1999}, and
in Ref.~\cite{Ga:2011} a new experimental technique called the {\it
identity method} was proposed. It solved the misidentification
problem for one specific combination of the second moments in a
system of two hadron species (`kaons' and `pions').
In Refs.~\cite{G:2011,RG:2012} this method was extended to show that
all the second moments as well as the higher moments of the joint
multiplicity distribution of particles of different types
can be uniquely reconstructed in spite of the effects of incomplete
identification. Notably, the results~\cite{G:2011,RG:2012} can be used for an
arbitrary number $k\geq 2$ of hadron species.
It is assumed that particle
identification is achieved by measuring the particle mass $m$.
Since any measurement is of finite resolution, we deal with
continuous distributions of observed masses  denoted as $\rho_j
(m)$ and normalized as ($j=1,\ldots,k\geq 2$)
\eq{ \label{norm-rho-i} \int dm \,\rho_j (m) = \langle N_j
\rangle~.
}
Note that for experimental data the functions $\rho_j(m)$ for particles
of type $j$ are obtained
from the inclusive distribution of the $m$-values for all particles from
all collision events. The identity variables $w_j(m)$ are
defined as
\eq{\label{wi}
w_j(m)~\equiv~\frac{\rho_j(m)}{\rho(m)}~,~~~~~ \rho (m) \equiv
\sum_{i=1}^k\rho_i (m) ~.
}
Complete identification (CI) of particles corresponds to
distributions $\rho_j (m)$ which do not overlap. In this case,
$w_j = 0$ for all particle species $i\neq j$ and $w_j = 1$ for the
$j$th species. When the distributions $\rho_j (m)$ overlap,
$w_j(m)$ can take the value of any real number from $[0,1]$.
We introduce the quantities
 \eq{\label{Wj2-def}
W_j~\equiv~ \sum_{i=1}^{N(n)}w_j(m_i)~,
~~~~W_j^2~\equiv~\Big(\sum_{i=1}^{N(n)}w_j(m_i)\Big)^2~,~~~~W_pW_q~\equiv~
\Big(\sum_{i=1}^{N(n)}w_p(m_i)\Big)\times
\Big(\sum_{i=1}^{N(n)}w_q(m_i)\Big)~,
}
with $j=1,\cdots,k$ and $1\le p<q\le k$,
and define their event averages  as
\eq{
  \langle W_j^2 \rangle~ =~
\frac{1}{N_{\rm ev}} \sum_{n=1}^{N_{\rm ev}} W_j^2~,~~~~
\langle W_pW_q \rangle ~=~ \frac{1}{N_{\rm ev}} \sum_{n=1}^{N_{\rm
ev}} W_pW_q~,
\label{Wj2-av}
}
where $N_{\rm ev}$ is the number of events, and
$N(n)=N_1(n)+\cdots+N_k(n)$ is the total multiplicity in the $n$th
event. Each experimental event is characterized by a set of
particle masses $\{m_1,m_2,\ldots,m_N\}$, for which one can
calculate the full set of identity variables:
$\{w_j(m_1),w_j(m_2),\ldots,w_j(m_N)\}$, with $j=1,\ldots,k$.
Thus, the quantities $W_j$, $W_j^2$, and $W_{p}W_q$ are completely defined
for each event, and their average values (\ref{Wj2-av}) can be
found experimentally by straightforward e-by-e averaging.
The main idea is to find the relations between these $W$-quantities
and the unknown moments of the multiplicity distribution $\langle N_j^2\rangle$
and $\langle N_pN_q\rangle$.
In the
case of CI, one finds $W_j=N_j$,
thus,
Eq.~(\ref{Wj2-av}) yields
\eq{\label{WW}
\langle W_j^2\rangle~=~\langle N_j^2\rangle~,~~~~\langle
W_pW_q\rangle~=~\langle N_pN_q\rangle~.
}
\subsection{Second Moments of Chemical Fluctuations}\label{ss-smcf}
The quantities $\langle W_j^2 \rangle$ and $\langle W_q
W_p\rangle$ can be calculated as follows \cite{G:2011}
\eq{
& \langle W_j^2\rangle ~=~
\sum_{i=1}^k\langle N_i \rangle \big[
u_{ji}^2~-~(u_{ji})^2\big]
 + \sum_{i=1}^k\langle N_i^2\rangle (u_{ji})^2
+ 2\sum_{1\leq i<l\leq k}\langle N_i N_l \rangle
u_{ji} u_{jl}~,\label{Wj2} \\
& \langle W_p W_q\rangle ~=~
\sum_{i=1}^k\langle N_i \rangle \Big[u_{pqi}
~ -~u_{pi} u_{qi}\Big]
 ~+ ~\sum_{i=1}^k\langle N_i^2\rangle u_{pi}
 u_{ki}
 ~ + ~\sum_{1\leq i<l\leq k}\langle N_i N_l \rangle
\Big[u_{pi} u_{ql}~ +~ u_{pl} u_{qi}\Big]~.\label{WpWq}
}
In Eqs.~(\ref{Wj2}) and (\ref{WpWq}), ${\cal P}(N_1,\ldots ,N_k)$
is the multiplicity distribution, $P_i (m) \equiv \rho_i(m)/
\langle N_i \rangle $ are the mass probability distributions of
the $i$th species, and ($s=1,2$)
\eq{\label{j-av}
u_{ji}^s ~\equiv~ \frac{1}{\langle N_i \rangle } \int dm \,
w_j^s(m)~ \rho_i (m)  ~,~~~~ u_{pqi}~\equiv~\frac{1}{\langle
N_i\rangle} \int dm\, w_p(m)\, w_q(m)~\rho_i(m)~.
}

In the case of CI,  when the distributions $\rho_j (m)$ do not
overlap, one finds that
\eq{\label{CI-j}
u_{ji}^{s} ~ =~\delta_{ji}~,~~~~~ u_{pqi}~ =~ 0~,
}
and Eqs.~(\ref{Wj2}) and (\ref{WpWq}) reduce then to
Eq.~(\ref{WW}). The incomplete particle identification transforms
the second moments $\langle N_j^2\rangle$ and $\langle
N_pN_q\rangle$ to the quantities $\langle W_j^2\rangle$ and
$\langle W_p W_q\rangle$, respectively. Each of the later
quantities contains  linear combinations of all the first and
second moments, $\langle N_i\rangle$ and $\langle N_i^2\rangle$,
as well as all the correlation terms $\langle N_i N_l \rangle$.
Having introduced the notations
\eq{
\langle W_j^2 \rangle  - \sum_{i=1}^k\langle N_i \rangle \big[
u_{ji}^2~-~ (u_{ji})^2\big] ~\equiv~b_j~,
~~~~ \langle W_pW_q\rangle -\sum_{i=1}^k\langle N_i \rangle \big[
u_{pqi}~ -~ u_{pi} u_{qi} \big] ~\equiv~b_{pq}~, \label{bpq}
}
one can transform Eqs.~(\ref{Wj2}) and (\ref{WpWq}) to the
following form:
\eq{
&\sum_{i=1}^k\langle N_i^2\rangle~ u_{ji}^2
+ 2\sum_{1\leq i<l\leq k}\langle N_i N_l \rangle ~ u_{ji} u_{jl}
~=~b_j~,~~~~j=1,2,\ldots,k~,\label{bj1}\\
& \sum_{i=1}^k\langle N_i^2\rangle ~u_{pi} u_{qi}
+ \sum_{1\leq i<l\leq k}\langle N_i N_l \rangle
\Big(u_{pi}u_{ql}~+~ u_{pl} u_{qi}\Big)~=~b_{pq}~,~~~~ 1\leq
p<q\leq k~.\label{bpq1}
}
The right-hand side of Eqs.~(\ref{bj1}) and (\ref{bpq1}) defined
by Eq.~(\ref{bpq}) are experimentally measurable quantities. The
same is true for the coefficients  $u_{ji}^{s}$ (with $s=1$ and
$2$) entering the left-hand side of Eqs.~(\ref{bj1}) and
(\ref{bpq1}).
Therefore, Eqs.~(\ref{bj1}) and (\ref{bpq1}) represent a system of
$k+ k(k-1)/2$ linear equations for the $k$ second moments $\langle
N_j^2\rangle$ with $j=1,\ldots,k$ and $k(k-1)/2$ correlators
$\langle N_pN_q\rangle$ with $1\leq p< q\leq k$.
In order to solve Eqs.~(\ref{bj1}) and (\ref{bpq1}) we introduce
the $[k+k(k-1)/2]\times[k+k(k-1)/2]$ matrix $A$
\eq{\label{AA}
A~=~
\begin{pmatrix}
a_{1}^{1} & \ldots & a_{1}^{k} &|&
a_{1}^{12}
&\ldots  & a_{1}^{(k-1)k}\\
. & . & . &|&
 . &  .  & .\\
. & . & . &|&
 . &  .  & .\\
 a_{k}^{1} & \ldots & a_{k}^{k} & | &
 a_{k}^{12}
 &\ldots &  a_k^{(k-1)k}  \\
--- & --- &--- &|&---& ---&---
\\
a_{12}^1 &\ldots  & a_{12}^k  & | & a_{12}^{12}
& \ldots & a_{12}^{(k-1)k}\\
. & . & . &|&
. & . & .  \\
. & . & . &|&
 . & . & .  \\
a_{12}^{k}& \ldots & a_{(k-1)k}^k  & | &   a_{(k-1)k}^{12} &
\ldots & a_{(k-1)k}^{(k-1)k}
 \end{pmatrix}~,
}
where
\eq{
a^{i}_{j}~&\equiv ~ u_{ji}^2~,~~1\leq i,j\leq k~;~~~~~ a_{i}^{pq}
\equiv ~2 u_{ip}u_{iq}~,~~ 1\leq p< q\leq k~,~~
i=1,\ldots ,k ~~;\\
a_{pq}^{i}~&\equiv ~u_{pi}u_{qi}~,~~1\leq p< q\leq k~,~~i=1,\ldots
,k~;
\\
 a_{pq}^{lm}~&\equiv~u_{pl}u_{qm}+u_{ql}u_{pm}~,~~
1\leq p<q\leq k~,~~1\leq l<m\leq k~.
}
The solution of Eqs.~(\ref{bj1}) and (\ref{bpq1}) can be presented
by Cramer's formulas in terms of the determinants
\eq{\label{Nj2-sol}
\langle N_j^2\rangle ~=~ \frac{{\rm det}~A_j}{{\rm
det}~A}~,~~~~~\langle N_pN_q\rangle ~=~ \frac{{\rm
det}~A_{pq}}{{\rm det}~A}~,~
}
where the matrices $A_j$ and $A_{pq}$ are obtained by substituting
in the matrix $A$ the column $ a_{1}^{j},\ldots, a_{k}^{j},
a_{12}^j,$ $\ldots, a_{(k-1)k}^j$ and the column $a_1^{pq},\ldots ,
a_k^{pq}, a^{pq}_{12},\ldots , a_{(k-1)k}^{pq}$, respectively, for
the column $b_1,\ldots, b_k,$ $b_{12},$ $ \ldots, b_{(k-1)k}$.
Therefore, if ${\rm det}A\neq 0$, the system of linear equations
(\ref{bj1}) and (\ref{bpq1}) has a unique solution (\ref{Nj2-sol})
for all the second moments.
In the case of CI (\ref{CI-j}), one finds ${\rm det}A=1$, ${\rm
det}A_j=b_j$,  and ${\rm det}A_{pq}=b_{pq}$. The solution
(\ref{Nj2-sol}) reduces then to Eq.~(\ref{WW}).

Introducing the  $[k+k(k-1)/2]$-vectors
\eq{\label{vectors}
{\cal N}~\equiv~
\begin{pmatrix}
\langle N_1^2\rangle  \\
\ldots \\
\langle N_k^2\rangle \\
\langle N_1N_2\rangle \\
\ldots\\
\langle N_{k-1}N_k \rangle
\end{pmatrix}
~,~~~~~~
 {\cal B}~\equiv~
\begin{pmatrix}
b_1  \\
\ldots \\
b_k \\
b_{12} \\
\ldots\\
b_{(k-1)k}
\end{pmatrix}~,
}
one can write Eqs.~(\ref{bj1}) and (\ref{bpq1}) in the matrix form
$A{\cal N} ={\cal B}$. The solution (\ref{Nj2-sol}) can be then
rewritten as
\eq{\label{AN}
{\cal N}~=~A^{-1}~{\cal B}~,
}
where $A^{-1}$ is the inverse matrix of $A$.
For two particle species, $k=2$, this solution takes the form
\eq{\label{A2}
\begin{pmatrix}
\langle N_1^2 \rangle \\
\langle N_2^2 \rangle \\
\langle N_1 N_2 \rangle
\end{pmatrix}
~=~
\begin{pmatrix} u_{11}^2 ~~~~& u_{12}^2~~~~& 2 u_{11} u_{12} \\
 u_{21}^2~~~~& u_{22}^2~~~~
&2 u_{21} u_{22} \\
u_{11} u_{21} ~~~~~ & u_{12} u_{22}~~~~ & u_{11} u_{22} + u_{12}
u_{21}
\end{pmatrix} ^{-1}~
\begin{pmatrix}
b_1\\
b_2\\
b_{12}
\end{pmatrix}
~.
}
Then Eq.~(\ref{A2}) yields
\eq{
 \langle N_1^2\rangle
~&=~ \frac{b_1u_{22}^2~+~b_2u_{12}^2~-~2b_{12}
u_{12}u_{22}}{\big(u_{11}u_{22}~-~u_{12}u_{21}\big)^2}~,
\label{N1-sol}
\\
 \langle N_2^2\rangle ~& = ~
 \frac{b_2u_{11}^2~+~
 b_1u_{21}^2 ~-~ 2b_{12}u_{21}u_{11}}
 {\big( u_{11}u_{22}~-~u_{12}u_{21}\big)^2}~,\label{N2-sol}\\
\langle N_1N_2\rangle ~& = ~
\frac{b_{12}\big(u_{11}u_{22}+u_{12}u_{21}\big) -b_1u_{22}u_{21}-
b_2u_{11}u_{12}}{\big(u_{11}u_{22}~-~u_{12}u_{21}\big)^2}~.
\label{N12-sol}
}
The above procedure eliminates the effect of misidentification and
provides the values of all the second moments $\langle
N_j^2\rangle$ and $\langle N_pN_q\rangle$ in a model-independent
way, as they would be obtained in an experiment in which each
particle is uniquely identified.
%
This method was generalized
to determine third and higher moments of the
multiplicity distributions in events consisting of an arbitrary
number of different particle species in Ref.~\cite{RG:2012}.
Measurements of the third and higher moments of
e-by-e fluctuations are expected to be more
sensitive for the search of the CP in A+A collisions \cite{cp1,cp2}.
First results on fluctuations based on the identity method
were presented in Refs.~\cite{ismd2013,NA49_identity,Mackowiak_CPOD2013}.

\section{Fluctuations at the Critical Point}\label{s-CP}
In this section we consider
the  van der Waals (VDW)
equation of state
with both repulsive ($b>0$) and attractive ($a>0$) terms
(see, e.g., Refs.~\cite{greiner,LL}):
\eq{\label{vdw-p-n}
p(V,T,N)~=~\frac{NT}{V~-~b\,N}~-~a\,\frac{N^2}{V^2}~=~\frac{n\,T}{1~-~bn}~-~a\,n^2~.
}
%
%
It gives an example of a  system with the CP.
The equation
of state (\ref{vdw-p-n}) was suggested in 1873,
and
for his work van der Waals obtained the Nobel Prize in physics in 1910.
The particle
proper volume
parameter in Eq.~(\ref{vdw-p-n}) equals to $b = 4\cdot (4\pi r^3/3)$
with $r$ being the
corresponding hard
sphere radius of particle.
The $b$ parameter describes the repulsion between particles and
can be rigorously obtained (in particular,
a factor of 4 in the
expression for $b$)
for a gas of the hard balls at low density (see, e.g., Ref. \cite{LL}).

\subsection{Phase Diagram}\label{ss-PD}
The VDW equation of state contains a 1$^{\rm st}$ order liquid-gas phase
transition and CP.
The thermodynamical quantities at the CP are equal to \cite{LL}:
\eq{\label{crit}
 T_c = \frac{8a}{27b}~,~~~~~ n_c =
\frac{1}{3b}~,~~~~~ p_c = \frac{a}{27b^2}~.
}

The VDW equation (\ref{vdw-p-n}) can be then rewritten in the following
dimensionless (reduced) form:
\begin{equation}
\widetilde{p}\ =\ \frac{8\,\widetilde{T}\, \widetilde{n}}{3 -
\widetilde{n}}\, -\, 3\,\widetilde{n}^2 \,, \label{vdw-dim}
\end{equation}
where $\widetilde{n} = n/n_c$, $\widetilde{p} = p/p_c$, and
$\widetilde{T} = T/T_c$.
In the dimensionless presentation (\ref{vdw-dim}) the VDW equation
has a  universal form independent of the values of $a$ and $b$, and
the CP
(\ref{crit}) is transformed to
$\widetilde{T}_c \, =\, \widetilde{p}_c\, =\,
\widetilde{n}_c\, =\, 1 \,.$

\begin{figure}[t]
\begin{minipage}{.49\textwidth}
\centering
\includegraphics[width=\textwidth]{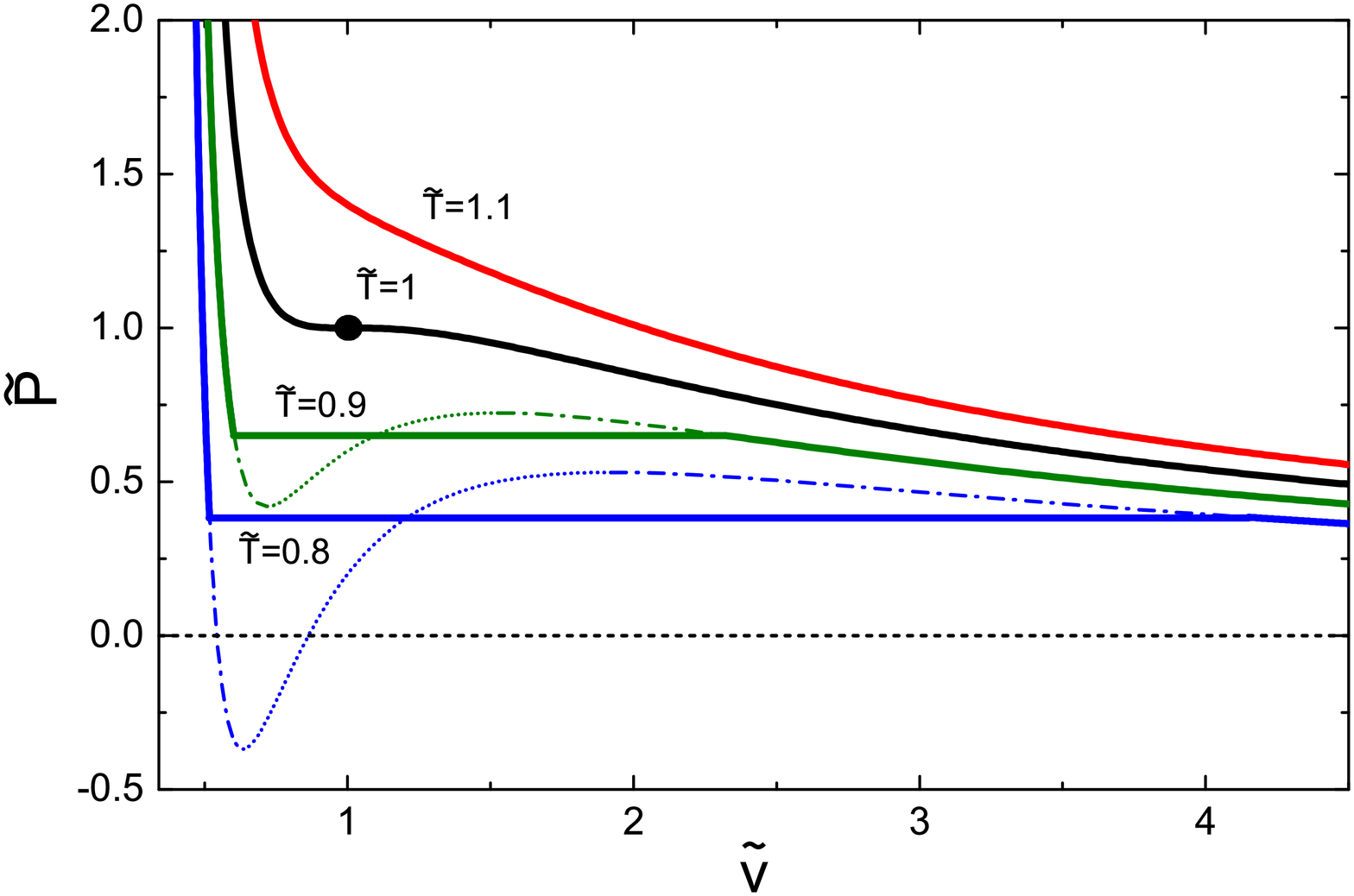}
\end{minipage}
\begin{minipage}{.49\textwidth}
\centering
\includegraphics[width=\textwidth]{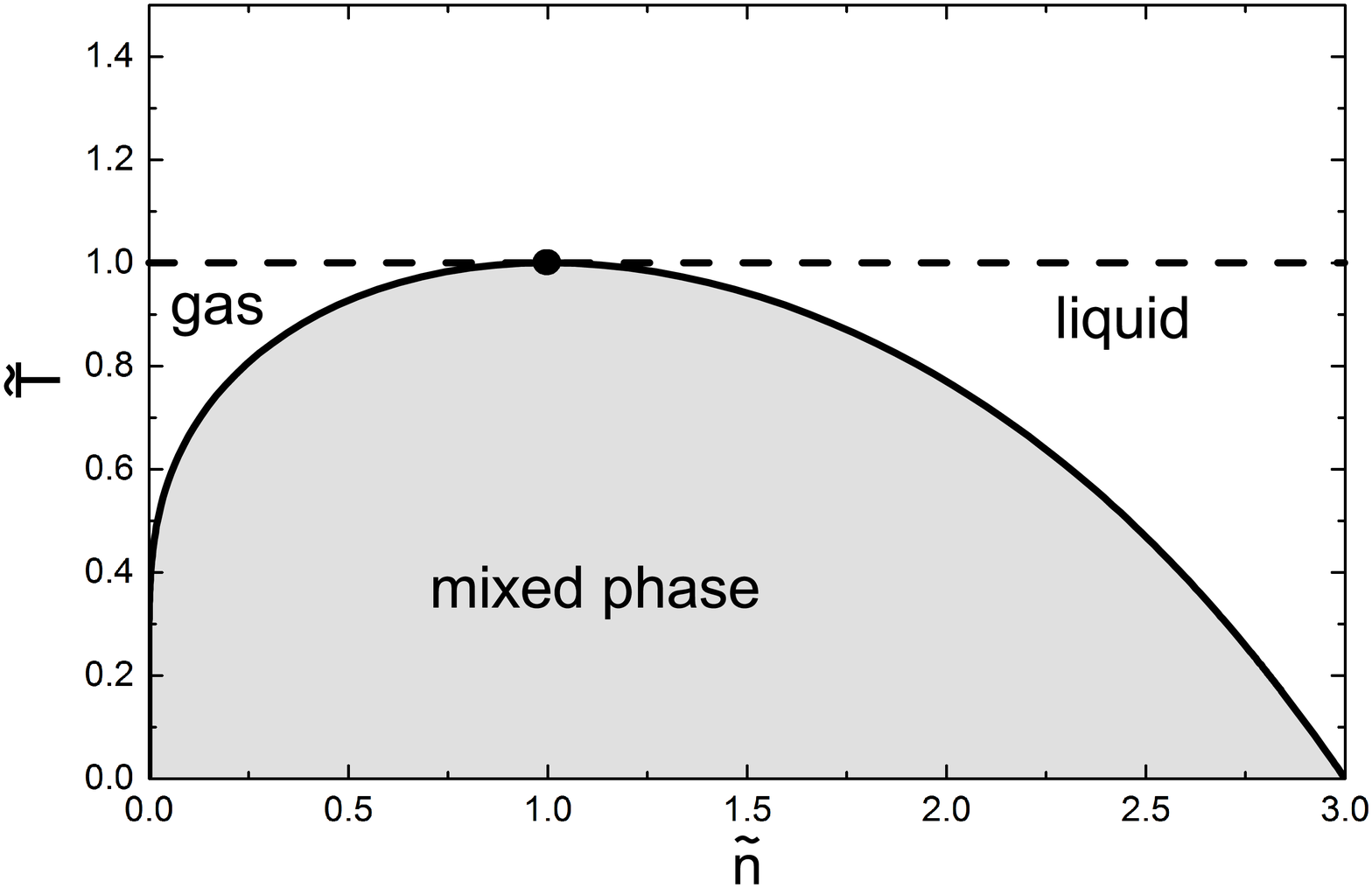}
\end{minipage}
\caption{{\it Left}: The dimensionless form of the VDW
isotherms for pressure, $\widetilde{p}$,
versus the volume per particle
$\widetilde{v}=\widetilde{n}^{-1}$.
The dashed-dotted lines
present the metastable parts of the VDW
isotherms at $\widetilde{T}<1$, whereas
the dotted lines correspond to unstable parts.  The full circle
on the $\widetilde{T}=1$ isotherm corresponds to the CP.
{\it Right}: The phase diagram for the VDW equation of state
on the $(\widetilde{n},\widetilde{T})$ plane.
The phase coexistence region
resulting from the Maxwell construction is depicted by grey shaded
area.
 } \label{fig:dim-isotherms}
\end{figure}

The dimensionless VDW isotherms
are presented in
Fig.~\ref{fig:dim-isotherms} ({\it left}) as functions of $\widetilde{v}\equiv \widetilde{n}^{-1}$ .
To describe the phase coexistence
region below the critical temperature the VDW isotherms
should be corrected by the well known Maxwell construction of {\it equal areas}.
These corrected parts of the VDW isotherms are shown by the solid horizontal lines.
Figure \ref{fig:dim-isotherms} ({\it right})
depicts the liquid-gas  coexistence region on
the $(\widetilde{n},\widetilde{T})$ plane.

\subsection{The GCE Formulation and Particle Number Fluctuations}
As the first step, the VDW equation (\ref{vdw-p-n}) will be
transformed to the GCE.
The GCE formulation will be then used to calculate the particle number
fluctuations in the VDW gas. These calculations were done
in Ref.~\cite{VAG:2015}, where details can be found.
The GCE form of the VDW equation of state has its simplest form
for the particle number density,
\eq{\label{nvdw}
n(T,\mu)~=~\frac{n_{\rm id}(T,\mu^*)}{1~+~b\,n_{\rm id}(T,\mu^*)}~,
~~~~~~\mu^*~=~
\mu~-~b\,\frac{n\,T}{1-b\,n}~+~2a\,n~,
}
as a function of temperature $T$ and chemical potential $\mu$.
If $a=b=0$, the function (\ref{nvdw}) is reduced to the ideal gas
particle number density. Note that $\mu$ regulates $\langle N\rangle_{\rm gce}$,
i.e., $N$ plays the role of conserved charge, and no antiparticles are introduced.
For the ideal Boltzmann gas one obtains
\eq{\label{nid}
n_{\rm id}(T,\mu)~=~\exp\left(\frac{\mu}{T}\right)\,\frac{g\,m^2\, T}{2 \pi^2} \, K_2\left(\frac{m}{T}\right)~,
}
where $g$ is the degeneracy factor,
$m$ is the particle mass, and $K_2$ is the modified Hankel function.
The function $n(T,\mu)$  is the solution of transcendental
equation (\ref{nvdw}).
The GCE pressure $p(T,\mu)$ for the VDW equation of state
is obtained substituting $n$ in Eq.~(\ref{vdw-p-n}) by function $n(T,\mu)$ (\ref{nvdw}).

The variance of the total particle number in the GCE can be
calculated as
\eq{\label{Var} Var[N]~\equiv~\langle N^2 \rangle~-~\langle
N\rangle^2~ =~ T \left(\frac{\partial{\langle N \rangle}}{\partial
\mu}\right)_{T,V} = T \, V \, \left(\frac{\partial{n}}{\partial
\mu}\right)_{T},
}
where symbol $\langle ...\rangle$ denotes the GCE averaging, and
$n(T,\mu)$ is the particle number density in the GCE. The scaled
variance for the particle number fluctuations is then \cite{VAG:2015}:
\eq{\label{omega-vdw}
\omega[N]~\equiv~\frac{Var[N]}{\langle
N\rangle}~ =~ \frac{T}{n} \, \left(\frac{\partial{n}}{\partial
\mu}\right)_{T}~=~
\left[\frac{1}{(1-bn)^2} - \frac{2an}{T}
\right]^{-1}~.
}
It is clearly seen from Eq.~(\ref{omega-vdw}) that in the VDW
gas the repulsive interactions suppress the particle number
fluctuations, whereas the attractive interactions lead to their
enhancement. Note that for $a=0$ the scaled variance
(\ref{omega-vdw}) is reduced to the result  for the excluded volume model
obtained earlier in Ref.~\cite{GHN}.

\begin{figure}[t]
\begin{minipage}{.49\textwidth}
\centering
\includegraphics[width=\textwidth]{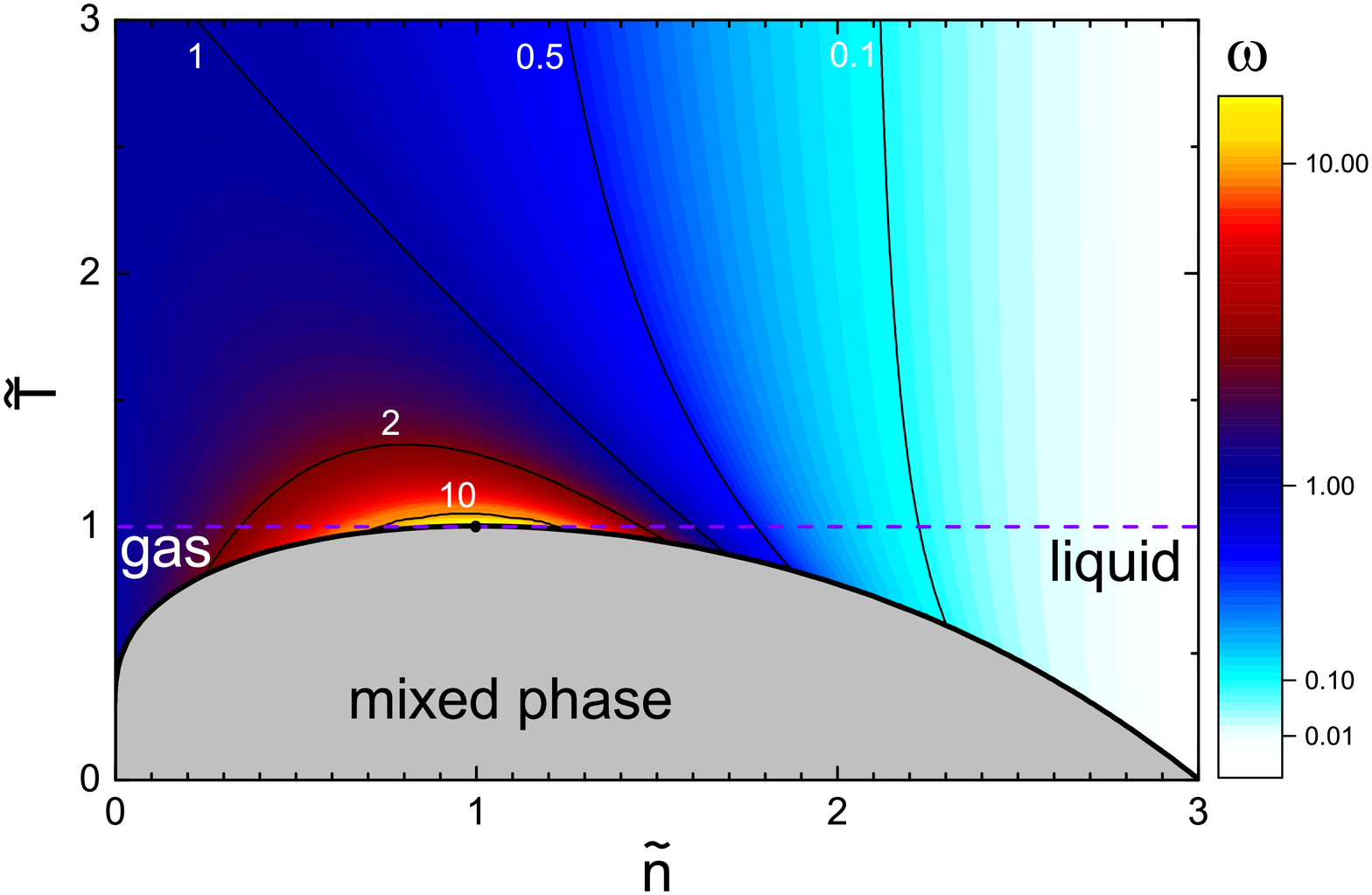}
\end{minipage}
\begin{minipage}{.49\textwidth}
\centering
\includegraphics[width=\textwidth]{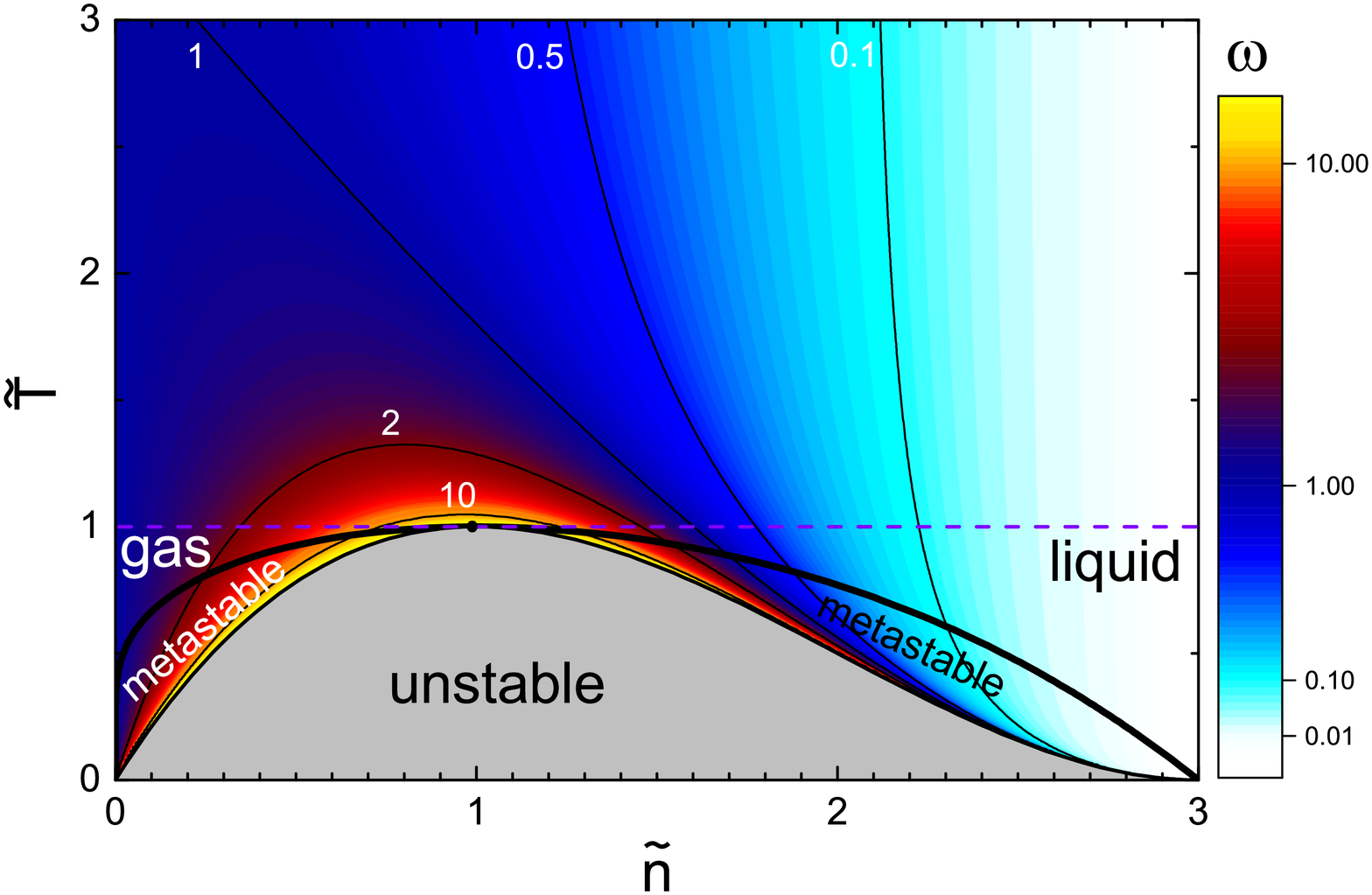}
\end{minipage}
\caption{{\it Left}:
The lines of constant values of the scaled variance $\omega[N]$ are shown
on the $(\widetilde{n},\widetilde{T})$ phase diagram,
outside of the mixed phase region.
{\it Right}: The lines of constant values of the scaled variance $\omega[N]$ are shown
on the $(\widetilde{n},\widetilde{T})$ phase diagram for
both stable and metastable pure phases.
The boundary between stable and metastable phases is depicted by the thick
black line, and the unstable region is depicted by the gray area.
} \label{fig-fluct}
\end{figure}

The scaled variance (\ref{omega-vdw}) expressed in terms $\widetilde{n}$ and
$\widetilde{T}$ equals to
\eq{\label{omega-2}
\omega[N]~ =~ \frac{1}{9}\,\left[\frac{1}{(3~-~\widetilde{n})^2}
~-~ \frac{\widetilde{n}}{4\,\widetilde{T}} \right]^{-1}~.
}
In Fig.~\ref{fig-fluct} the lines of constant values of $\omega[N]$ are shown
on the $(\widetilde{n},\widetilde{T})$ phase diagram outside of the mixed phase region.
\begin{figure}[t]
\begin{minipage}{.98\textwidth}
\centering
\includegraphics[width=\textwidth]{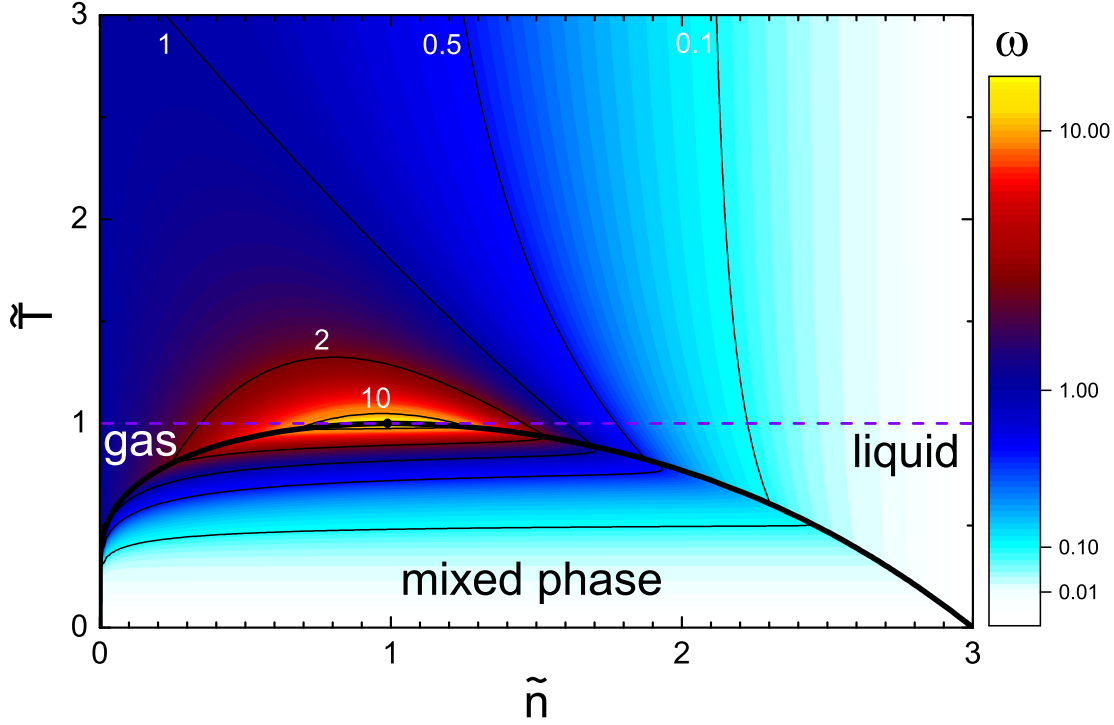}
\end{minipage}
\caption{
The lines of constant values of $\omega[N]$ are shown
on the  phase diagram
for $0<\tilde{T}<3$
and for
all possible $\widetilde{n}$ values, for stable states
both inside and outside the mixed phase region.
 } \label{fig-fluct-both}
\end{figure}
At any fixed value of $\widetilde{T}$, the particle number fluctuations (\ref{omega-2})
approach to those of the ideal gas, i.e.,  $\omega[N]\cong 1$,
at $\widetilde{n}\rightarrow 0$,
and become small, $\omega[N]\ll 1$, at $\widetilde{n}\rightarrow 3$.
As it should be, the scaled variance (\ref{omega-2}) is always positive for all
possible values of $\widetilde{n}$ and $\widetilde{T}$ outside of the mixed phase region.
At $\widetilde{T}\rightarrow 0$ this is insured by simultaneous limiting behavior
of $\widetilde{n}\rightarrow 0$ in the pure gaseous phase and $\widetilde{n}\rightarrow 3$
in the pure liquid phase.

\subsection{Critical Point}

At the CP ($\widetilde{T}=\widetilde{n}=1$)
the scaled variance of the particle number fluctuations diverges in the  GCE.
To study the behavior of $\omega[N]$ in a vicinity of the critical point
we introduce the quantities $\tau =\widetilde{T}-1 \ll 1$
and $\rho=\widetilde{n}-1\ll 1$.
Expanding (\ref{omega-2}) at small $\tau$
and $\rho$ and keeping only the lowest orders
of them one finds
\eq{\label{omega-c}
\omega[N]~\cong ~\frac{4}{9}\,\left[ \tau~+~ \frac{3}{4}\rho^2~+~ \tau\,\rho\right]^{-1}~.
}
In particular,
\eq{\label{omega1-c}
\omega[N]~\cong~ \frac{4}{9}\,\tau^{-1}~~ {\rm at}~~\rho=0~,
~~~~~~~~~ {\rm and}~~\omega[N]~\cong~ \frac{16}{27}\,\rho^{-2}~~
{\rm at}~~ \tau=0~.
}
Note that thermodynamical parameters
$\widetilde{T}$ and $\widetilde{n}$ correspond to
points outside the mixed phase region. This is shown
in Fig.~\ref{fig-fluct} {\it left}. Thus,
in Eqs.~(\ref{omega-c}) and (\ref{omega1-c}), parameter $\tau$ is positive,
while $\rho$ can be both positive and negative.

The VDW equation of state permits the existence
of metastable phases of super-heated liquid and
super-cooled gas. These states are depicted by the
dash-dotted lines on the VDW isotherms in Fig.~\ref{fig:dim-isotherms} ({\it left}).
In metastable phases the system is assumed to be uniform
and, therefore, one can use Eq.~(\ref{omega-2})
to calculate particle number fluctuations in
these phases.

The lines of constant values of $\omega[N]$ are shown
on the $(\widetilde{n},\widetilde{T})$ phase diagram in Fig.~\ref{fig-fluct}
{\it right} for both stable and metastable pure phases,
while the unstable region is depicted by the gray area.
It is seen that the scaled variance remains finite,
and diverges only at the boundary between the metastable and unstable regions.
We recall that at this boundary $\partial \widetilde{p} / \partial \widetilde{n} = 0$,
where $\widetilde{p}$ is the dimensionless CE pressure~\eqref{vdw-dim}. One can easily show
using Eqs.~\eqref{vdw-dim} and \eqref{omega-2} that $\omega[N] \to \infty$
when $\partial \widetilde{p} / \partial \widetilde{n} = 0$. Note that metastable
regions of the equation of state can be reached within fast non-equilibrium
processes, whereas the unstable region
is physically forbidden. Note also that the thermodynamical relations are
not fulfilled in the unstable region, e.g., nonphysical behavior with
$\omega[N] < 0$ is found in this region.

Finally, the scaled variance $\omega[N]$ is shown in Fig.~\ref{fig-fluct-both}
for all possible values of $\widetilde{n}$ and $\widetilde{T}$, both inside and
outside of the mixed phase. Now only thermodynamically stable states are considered.
Details of the calculations inside the mixed phase are presented in Ref.~\cite{VAG:2015}.
In a vicinity of the critical point inside the mixed phase we introduce
$0<t=1-\widetilde{T}\ll 1$ and  find
at $t\rightarrow 0$,
\eq{\label{omegaNc}
\omega[N]~\cong~\frac{16}{9} ~t^{-1}~.
}
Thus, the scaled variance $\omega[N]$
diverges at the critical point reached
from both outside and
inside the
mixed phase.

\section{Summary}\label{sec-sum}

1. The global conservation of energy and conserved charges influences the measured
e-by-e fluctuations of all observables. Thus, to calculate these fluctuations within statistical
mechanics one should properly choose the statistical ensemble. The GCE, CE, and MCE
are only some particular examples. Real situation may correspond to very different externally given
distributions of the volume $V$, energy $E$, and conserved charge(s) $Q$.

2. To avoid the trivial contributions from the system size fluctuations the strongly intensive
measures of e-by-e fluctuations should be used. These $\Delta[A,B]$ and $\Sigma[A,B]$
measures are constructed from
the second moments $\langle A^2\rangle$,  $\langle B^2\rangle$, and $\langle AB\rangle$
of two extensive quantities $A$ and $B$.

3. A special procedure, the identity method, should be used to calculate the chemical fluctuations
in a case of the incomplete particle identification.
It solves the misidentification problem for the second and higher moments.
Thus the  joint multiplicity distribution of particles of different types
(a number of species $k\ge 2$) can  be uniquely reconstructed.

4. The van der Waals equation of state is considered
in the GCE, and particle number fluctuations are calculated.
The scaled variance $\omega[N]$  diverges
at the the critical point.
Admitting a presence of metastable states, one observes that the scaled variance $\omega[N]$
diverges also at the boundary between the metastable and unstable regions, where
$\partial p / \partial n = 0$.

%
\vspace{0.5cm}
\noindent
{\bf Acknowledgements.}~~ This work was supported by the Humboldt
Foundation and by the Program of Fundamental Research
of the Department of Physics and Astronomy of National Academy of Sciences, Ukraine

\end{document}